\documentclass[letter]{aa}

\usepackage{xspace}
\usepackage{graphicx}
\usepackage{amssymb}
\usepackage{txfonts}
\usepackage{multicol}

\newcommand{\cc}{\ensuremath{{\rm cm}^{-3}}\xspace}
\newcommand{\amm}{\ensuremath{{\rm NH}_3}\xspace}
\newcommand{\tex}{\ensuremath{T_{\rm ex}}\xspace}
\newcommand{\tk}{\ensuremath{T_{\rm K}}\xspace}
\newcommand{\sig}{\ensuremath{\sigma_{\rm{v}}}\xspace}
\newcommand{\vel}{\ensuremath{\rm{v}_{LSR}}\xspace}
\newcommand{\htw}{\ensuremath{\rm H_2}\xspace}
\newcommand{\msun}{\ensuremath{M_\odot}\xspace}

\newcommand{\kms}{\ensuremath{{\rm km\,s}^{-1}}\xspace}
\newcommand{\ms}{\ensuremath{{\rm m\,s}^{-1}}\xspace}

\begin{document}

  \title{Ubiquitous \amm supersonic component in L1688 coherent cores}
  
   \author{
        Spandan Choudhury
          \inst{1}
          \and
          Jaime E. Pineda \inst{1}
          \and
          Paola Caselli \inst{1}
          \and
          Adam Ginsburg \inst{2}
          \and
          Stella S. R. Offner \inst{3}
          \and
          Erik Rosolowsky \inst{4}
          \and
          Rachel K. Friesen \inst{5}
          \and
          Felipe O. Alves \inst{1}
          \and
          Ana Chac\'on-Tanarro \inst{6}
          \and
          Anna Punanova \inst{7}
          \and
          Elena Redaelli \inst{1}
          \and
          Helen Kirk \inst{8,9}
          \and
          Philip C. Myers \inst{10}
          \and
          Peter G. Martin \inst{11}
          \and
          Yancy Shirley \inst{12}
          \and
          Michael Chun-Yuan Chen \inst{8}
          \and
          Alyssa A. Goodman \inst{10}
          \and
          James Di Francesco \inst{13}
          }

    \institute{
        Max-Planck-Institut f\"ur extraterrestrische Physik, Giessenbachstrasse 1, D-85748 Garching, Germany\\
        \email{spandan@mpe.mpg.de}
        \and
        Department of Astronomy, University of Florida, PO Box 112055, USA
        \and
        Department of Astronomy, The University of Texas at Austin, Austin, TX 78712, USA
        \and
        Department of Physics, 4-181 CCIS, University of Alberta, Edmonton, AB T6G 2E1, Canada
        \and
        Department of Astronomy \& Astrophysics, University of Toronto, 50 St. George St., Toronto, ON M5S 3H4, Canada
        \and
        Observatorio Astron\'omico Nacional (OAN-IGN), Alfonso XII 3, 28014, Madrid, Spain
        \and
        Ural Federal University, 620002 Mira st. 19, Yekaterinburg, Russia
        \and
        Department of Physics and Astronomy, University of Victoria, 3800 Finnerty Rd., Victoria, BC V8P 5C2, Canada
        \and
        Herzberg Astronomy and Astrophysics, National Research Council of Canada, 5071 West Saanich Rd., Victoria, BC V9E 2E7, Canada
        \and
        Center for Astrophysics, Harvard \& Smithsonian, 60 Garden Street, Cambridge, MA 02138, USA
        \and
        Canadian Institute for Theoretical Astrophysics, University of Toronto, 60 St. George St., Toronto, ON M5S 3H8, Canada
        \and
        Steward Observatory, 933 North Cherry Ave., Tucson, AZ 85721, USA
        \and
        Herzberg Astronomy \& Astrophysics Research Centre, National Research Council of Canada, 5071 West Saanich Road, Victoria, BC, V9E 2E7 Canada
        }
    \date{}

 
  \abstract
   {Star formation takes place in cold dense cores in molecular clouds. Earlier observations have found that dense cores exhibit subsonic non-thermal velocity dispersions. In contrast, CO observations show that the ambient large-scale cloud is warmer and has supersonic velocity dispersions.}
   {We aim to study the ammonia (\amm) molecular line profiles with exquisite sensitivity towards the coherent cores in L1688 in order to study their kinematical properties in unprecedented detail.}
   {We used \amm (1,1) and (2,2) data from the first data release (DR1) in the Green Bank Ammonia Survey (GAS). We first smoothed the data to a larger beam of 1\arcmin\ to obtain substantially more extended maps of velocity dispersion and kinetic temperature, compared to the DR1 maps. We then identified the coherent cores in the cloud and analysed the averaged line profiles towards the cores.}
   {For the first time, we detected a faint (mean \amm(1,1) peak brightness $<$0.25\,K in $T_{MB}$), supersonic component towards all the coherent cores in L1688. We fitted two components, one broad and one narrow, and derived the kinetic temperature and velocity dispersion of each component. 
   The broad components towards all cores have supersonic linewidths ($\mathcal{M}_S \ge 1$). 
   This component biases the estimate of the narrow dense core component's velocity dispersion by $\approx$28\% and the kinetic temperature by $\approx$10\%, on average, as compared to the results from single-component fits.}
   {Neglecting this ubiquitous presence of a broad component towards all coherent cores causes the typical single-component fit to overestimate the temperature and velocity dispersion. This affects the derived detailed physical structure and stability of the cores estimated from \amm observations.}


   \maketitle
%

\section{Introduction}

Star formation takes place in dense cores, which are embedded in molecular clouds. These cores in various molecular clouds are therefore studied in detail in relation to their physical and chemical properties. Across different molecular clouds, the star-forming cores are characterised by higher density and lower temperatures, as compared to the parental cloud \citep{myers_1983_SF_core,myers-benson_1983_dense_core,caselli_2002_dense_core}. When studied using molecular transitions tracing higher densities ($n(\htw) > 10^4$ \cc), the cores are revealed to show 
subsonic levels of turbulence \citep{coh_core_barranco_goodman_1998, kirk_2007_rel_vel,roso_2008_amm_dens_core}. This is in contrast to lower density gas surrounding the cores, which show supersonic linewidths \citep{coh_core_goodman_barranco_1998}.

Thanks to the numerous hyperfine components, \amm remains optically thin in the individual components even at high column densities \citep[see][]{caselli_2017_amm}, thereby making it an important and useful tracer of cold and dense gas.
\citet{coh_core_barranco_goodman_1998} studied four cores in \amm (1,1) emission, and found that the linewidths within the cores were roughly constant and slightly greater than the pure thermal value. They called these `coherent cores'.
A `transition to coherence' from the turbulent environment to the coherent region was hypothesised  \citep{coh_core_goodman_barranco_1998}, but not directly observed.
\citet{williams_2000_coher} also noted eight `quiescent' cores, which they defined as local minima in non-thermal dispersion, in the observation of Serpens in $\rm N_2H^+$, using BIMA (Berkeley Illinois Maryland Association). They showed that the dispersion increased outwards from the centres of these cores.
Using \amm (1,1) observations with the Green Bank Telescope, \citet{pineda2010} studied the transition from the surrounding gas to inside a core, for the first time in the same tracer in the B5 region in Perseus, and reported a sharp transition to coherence (a decrease in the velocity dispersion by a factor of 2 within 0.04 pc).
This {suggests that we can} define the boundaries of coherent cores in a systematic fashion, as regions with subsonic non-thermal linewidths \citep[see also][]{chen2019}. 
{It should be noted that this definition of cores does not necessarily define cores identically to those defined using continuum emission. Therefore, not all of the `coherent cores' have continuum counterparts.}

In the Green Bank Ammonia Survey \citep[GAS;][]{GASDR1}, star forming regions in the Gould Belt were observed using \amm hyperfine transitions. 
The first data release (DR1) includs the following four regions in nearby molecular clouds: B18 in Taurus, NGC 1333 in Perseus, L1688 in Ophiuchus, and Orion A North. Using the results from this survey and \htw column densities derived from Herschel data, \citet{chen2019} identified 12 coherent structures (termed `droplet' in the paper) in L1688 and six in B18.
They observed that these droplets show high gas density ($\rm \langle n_{\htw} \rangle \approx 5 \times 10^4$ cm$^{-3}$, derived from their masses and radii) and near-constant, almost-thermal velocity dispersions, with a sharp velocity dispersion increase across the boundary. 

This letter presents a new analysis of the GAS DR1 observations to reach unprecedented noise levels towards coherent cores and reveal a broad supersonic component that has been previously unidentified.
We focus on L1688, which is part of the Ophiuchus molecular cloud \citep[distance = 138.4$\pm$2.6~pc,][and mass $\sim$980~\msun, \citealp{ladj_2020_hgbs_oph_strc}]{l1688_dist_ortiz-leon} because it is one of the nearest star-forming regions and with the most extended \amm emission beyond cores among the GAS DR1 regions. 


\section{Ammonia maps}
\label{sec_amm_data}

We use the \amm (1,1) and (2,2) maps from GAS DR1 \citep[][]{GASDR1}. The observations were carried out using the Green Bank Telescope (GBT) to map \amm emission in the star-forming regions in the Gould Belt with $A_V>$7 mag using the seven-beam K-Band Focal Plane Array (KFPA) at the GBT. Observations were performed in frequency switching mode with a frequency throw of 4.11 MHz ($\approx$ 52 \kms at 23.7 GHz). The spectral resolution of the data is 5.7 kHz, which corresponds to $\approx$ 72.1 \ms at 23.7 GHz (i,e, the approximate frequency of observations). The extents of the maps were selected using continuum data from Herschel or JCMT, or extinction maps derived from 2MASS (Two Micron All Sky Survey). To convert the spectra from frequency to velocity space, the rest frequencies for \amm (1,1) and (2,2) lines were considered to be 23.6944955 GHz and 23.7226336 GHz, respectively.

The parameter maps of L1688, which were released in DR1, are very patchy, particularly for the kinetic temperature. Therefore, to obtain a good estimate of the Mach number throughout the cloud, which requires determining the temperature, and thus, a good detection of the (2,2) line, the data were smoothed by convolving them to a beam of 1\arcmin \ (GBT native beam at 23 GHz is $\approx$31$\arcsec$). The data cube was then re-gridded to avoid oversampling. The relative pixel size was kept the same as in the original GAS maps, at one-third the beam width. The median noise level achieved as a result is 39 mK in the regions of interest (see Sections \ref{sec_id_coh} and \ref{sec_spec_avg}). In comparison, the median noise in the same regions was 131 mK in the original GAS DR1 maps.\footnote{The GAS data use a tapered Bessel function as the gridding kernel \citep[advocated for by][]{mangum_2007_kernel}, to achieve maximum resolution. The kernel is not strictly positive, which leads to adjacent pixels having anti-correlated noise.  When smoothing over anti-correlated values, the noise level drops faster than what is expected from independent data.
Therefore, we see the noise level drop faster than the inverse of the beam radius.}

\section{Analysis}
\label{sec_analy}

\subsection{Line fitting}
\label{sec_line_fit}

We fitted \amm line profiles to the data with the \verb+pyspeckit+ package \citep{pyspeckit}, which uses a forward modelling approach, following the process described in \citet{GASDR1}. 
The range in velocity to fit is determined from the average spectrum of the entire region (Figure \ref{avg_spec_whole}). 
Since the ortho-to-para ratio {in the region is not known}, we only used the p-\amm column density in the fitting process, and did not attempt to convert it to total \amm column density\footnote{To compare N(p-NH3) to the total \amm column densities reported in other works, an easy conversion is to multiply it by two, with the ortho-to-para ratio assumed to be unity.}.

The model produces synthetic spectra based on initial guesses provided for the input parameters: excitation temperature (\tex), kinetic temperature (\tk), para-ammonia column density (N(p-\amm)), velocity dispersion (\sig), and line-of-sight central velocity (\vel) of the gas \citep[Section 3.1 in][]{GASDR1}. A non-linear gradient descent algorithm, MPFIT \citep{markwardt2009} is then used to determine the best-fit model, and the corresponding values of the parameters. A good set of initial conditions is necessary to ensure that the non-linear least-squares fitting does not get stuck in a local minimum. The value of \vel is critical, and, therefore, we used the first-order moments of the (1,1) line as initial guesses. The second-order moments were used as guesses for velocity dispersion. For the other parameters, we used guesses based on the GAS DR1 map of the respective parameter. The initial guesses used in our work are:
$\log_{10} (N_{p-\amm}/ {\rm cm^{-2}}) = 14$, $\tk = 20$\,K, and $\tex = 5$\,K. These numbers are within the range of the values reported in the GAS DR1 maps, and therefore, they are reasonable guesses. As a test, we checked the fits with varying guesses and determined that the exact values of the initial parameters did not affect the final results, as long as they were within a reasonable range.

In this work, we use the \verb+cold_ammonia+ model in the \verb+pyspeckit+ library, which makes the assumption that only the (1,1), (2,2) and (2,1) levels are occupied. It is also assumed that the excitation temperature, \tex, is the same for the (1,1) and (2,2) transitions, as well as their hyperfine components.

\subsection{Identification of coherent cores }
\label{sec_id_coh}

To study the spectra towards cores, we first define `coherent cores' as the region with 
a sonic Mach number $<1$ and larger than a beam  \citep[similar to][]{pineda2010,chen2019}\footnote{It should be noted that \citet{chen2019} also require a corresponding peak in N(\htw) inside a coherent region for it to be identified as a `droplet', whereas we consider the sonic Mach number as the only criterion.}. 
The sonic Mach number, $\mathcal{M_S}$, is defined as the ratio of the non-thermal velocity dispersion, $\sigma_{\rm NT}$ (see Appendix \ref{app_sig_nt_calc}) to the sound speed in the medium,
\begin{equation}
    \mathcal{M_S} = \frac{\sigma_{NT}}{c_S}~,
\end{equation}
where $c_S$ is the one-dimensional sound speed in the gas, 
\begin{equation}
    c_S = \sqrt{\frac{k_B T_K}{\mu_{gas}}}~.
\end{equation}
Here $\rm{k_B}$ is the Boltzmann's constant, $\rm{T_K}$ is the kinetic temperature in the region, and $\rm{\mu_{gas}}=2.37$ amu is the average molecular mass \citep{jens_2008_mu}.

Based on this definition, we identify 12 regions in L1688 as coherent cores. They are shown in Figure \ref{coh_cores}. See Appendix \ref{app_core_names} for a brief description of the cores.

\begin{figure*}[!t]  
\centering
\includegraphics[width=0.67\textwidth]{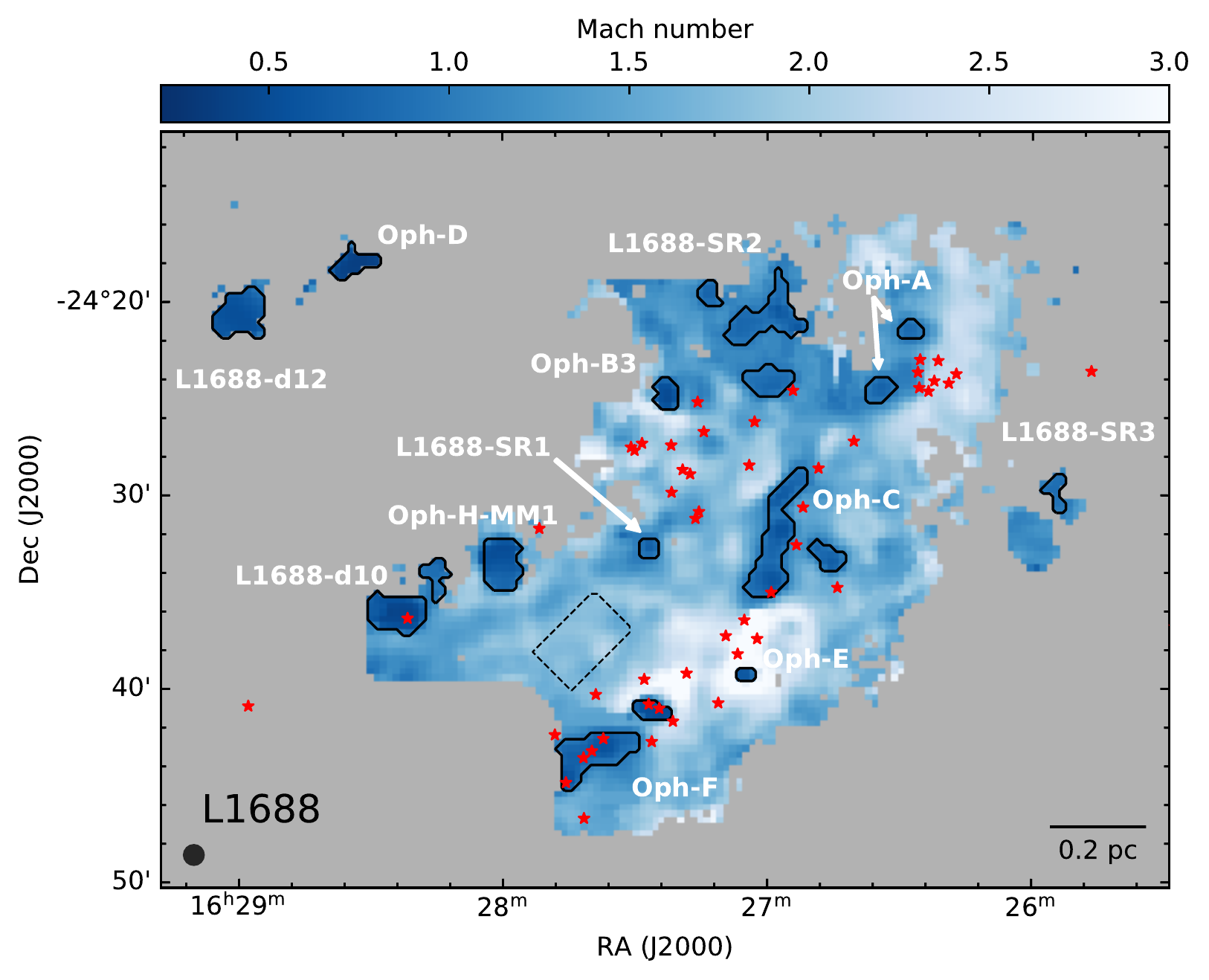}
\caption{Coherent cores, as defined in Section \ref{sec_id_coh}, are shown by black contours. The background colour scale shows the sonic Mach number. The dashed rectangle shows the box considered to be representative of the ambient cloud. 
The red stars show positions of Class 0/I and flat-spectrum protostars in the cloud \citep{yso_l1688}. The beam and the scale bar are shown in the bottom left and bottom right corners, respectively.} 
\label{coh_cores}
\end{figure*}

\subsection{Averaging spectra towards the coherent cores}
\label{sec_spec_avg}

We first averaged the spectra towards each coherent core defined in Section \ref{sec_id_coh}, to get a higher signal-to-noise ratio. To avoid any possible line broadening due to averaging in a region with velocity gradients, we aligned the spectra at each pixel within a core before averaging. We took the velocity at a pixel, which was determined from the single-component fit at that pixel, and using the \texttt{channelShift} function from module \texttt{gridregion} in the GAS pipeline\footnote{https://github.com/GBTAmmoniaSurvey/GAS/tree/master/GAS}, we shifted the spectra by the corresponding number of channels. Then, we averaged the resultant spectra from all pixels inside a core, which are now essentially aligned at v=0. 
This averaging allowed us to reach a typical noise level of 18 mK in the average spectra, which is almost seven times better than the median noise level of 131 mK in the individual spectra towards the cores in the original GAS maps.

We checked if smoothing the data (Section \ref{sec_amm_data}) before averaging had any effect on the results and found that the change is not significant. See Appendix \ref{app_chk_unsmth} for details.

\section{Results}
\label{sec_res}

\subsection{Presence of a second component in core spectra}
\label{sec_comp_1-2_comp}

The top panels in Figure \ref{avg_spec_ophF} show the average spectra in the Oph-F core, with a single-component fit (see Appendix \ref{app_avg_fit} for the guesses used in the fit process), and corresponding residual. Oph-F is shown here as an example, the average spectra for the other coherent cores are shown in Appendix \ref{app_avg_spec} (Figures \ref{avg_spec_ophA} to \ref{avg_spec_west}). From the residuals, it is evident that the fit does not recover all of the flux at the positions of the hyperfine lines. In particular, the  single-component fit misses the flux present in a broad component, which is ubiquitously present towards all cores.

Therefore, we added a second component to the fit (see Appendix \ref{app_avg_fit} for the fit procedure). 
The resultant two-component fits and the individual components for Oph-F are shown in the bottom panel of Figure \ref{avg_spec_ophF}, and in Figures \ref{avg_spec_ophA} to \ref{avg_spec_west} for the other cores. The two-component fit is clearly an improvement (better $\chi^2$), as seen from the respective residuals. 
We employed the Akaike Information Criterion (AIC, see Appendix \ref{app_mod_sel}) to verify that for each core, considering a second component corresponds to a significant improvement in the fit to the spectra. The difference in AIC values from single-component fits to two-component fits, $\rm \Delta_{AIC} = AIC_{1-comp.} - AIC_{2-comp.}$, are shown in Table \ref{tab_fit_para}. 

In model comparison, the model with the lowest AIC value is the preferred one, and, therefore all regions are better fit by a two-component fit. 
However, for coherent cores L1688-d12 and L1688-SR3, the value of $\rm \Delta_{AIC}$ is small (see Table \ref{tab_fit_para}) and  
the average spectra fits (Figures \ref{avg_spec_d12} and \ref{avg_spec_west}) only show marginal improvements with the two-component fit.
Therefore, we note that these two regions are better fit with a two-component model, but their fit-derived results are not as well-constrained as in the other ten coherent cores.

The kinetic temperatures, velocity dispersions, LSR velocities, and p-\amm column densities for the single-component fits and for each individual component in the two-component fits are shown in Table \ref{tab_fit_para}. We see that one of the two components is subsonic ($\mathcal{M}_S$< 1) across all cores, whereas the other component has $\mathcal{M}_S$ $\geq$1. 
We refer to the subsonic and supersonic components as `narrow' and `broad' components, respectively.

We note that the residuals from the two-component fit in some cores, in particular Oph-E, might suggest the presence of a third component or a non-Gaussian component. However, we do not see this towards all of the cores. Even when this occurs, the indication is not very clear : residuals comparable to the noise or not more than one or two channels wide. 
Therefore, we did not fit more than two components to the spectra in the cores.

\begin{figure*}[!ht]  
\centering
  \includegraphics[width=0.8\textwidth]{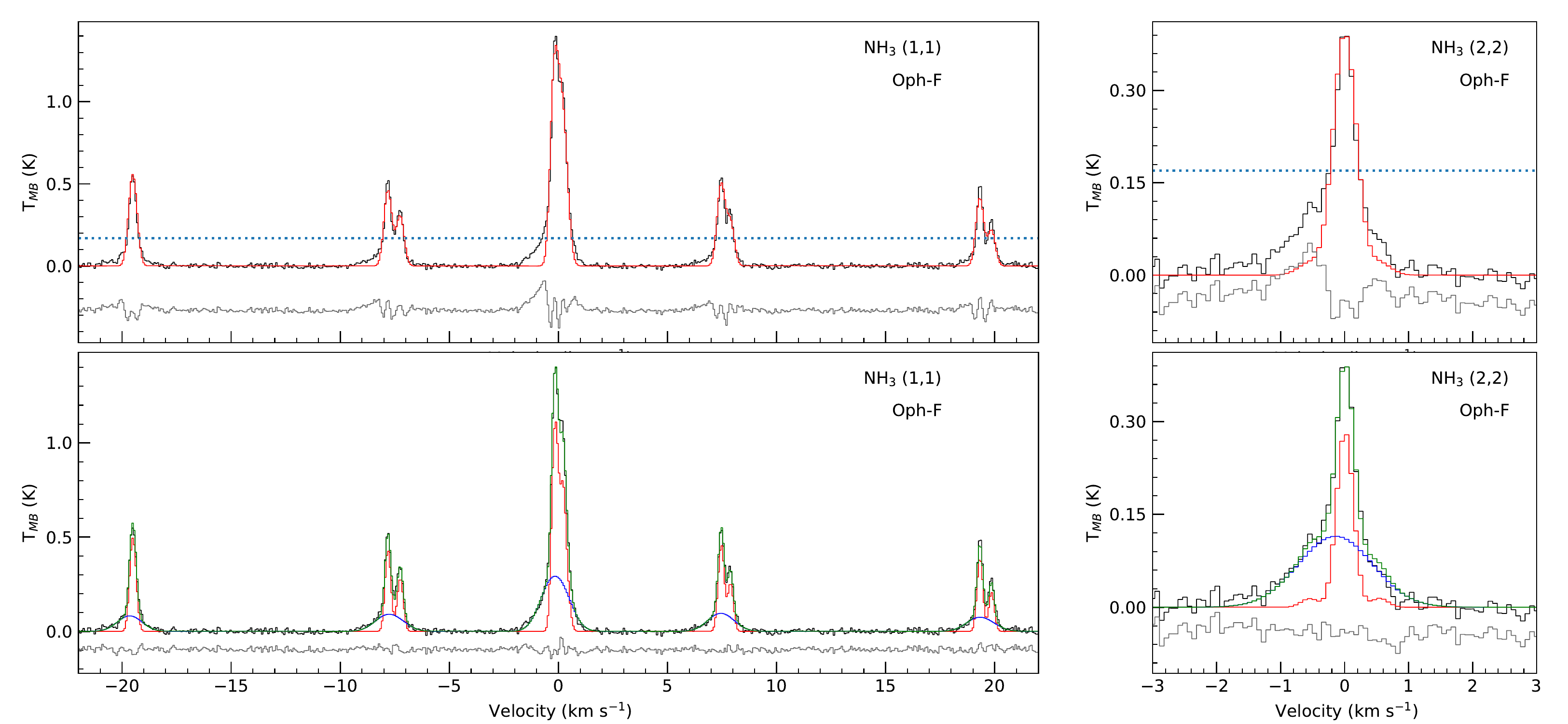}
  \caption{{\it Top panels} : Average \amm (1,1) and (2,2) spectra of Oph-F core, with a single-component fit. The blue-dotted line shows the median noise level in Oph-F in the original GAS data, which clearly shows that the broad component cannot be detected in the individual spectra in GAS DR1 data. {\it Bottom panels} : Same spectra, with two-component fit (green). The narrow (red) and broad (blue) components are also shown separately.}
     \label{avg_spec_ophF} 
\end{figure*}

\subsection{Detection of a broad component in cores}
\label{sec_dtct_brd}

A narrow and a broad component of ammonia emission has been detected in all of the coherent cores in L1688 (Figure \ref{coh_cores}). The broad component in the cores is fainter, with a mean (1,1) peak brightness temperature ($\rm{T_p}$) $\approx$0.25 K
and $\rm T_{p,(2,2)}$< 0.13 K, except in Oph-A ($\rm T_{p,(2,2)}$$\approx$0.27 K) and Oph-C ($\rm T_{p,(2,2)}$$\approx$0.16 K). Because of such low intensities, this component cannot be detected from the previous ammonia maps. As a comparison, the original GAS maps had noise levels of $\approx$131 mK in the coherent core regions, which is not low enough for a 3$\sigma$ detection of the (1,1), and in most cases, this is comparable to the \ensuremath{\rm T_{MB}} of the broad component in the (2,2). Therefore, using the original GAS data, one would not be able to detect the broad component, and the temperature and linewidth of the narrow core component would be affected. As can be seen in Table \ref{tab_fit_para}, the mean noise level in the average spectra in the cores is $\approx$18 mK in this work, enabling us to comfortably fit even the faint broad component in these spectra for both (1,1) and (2,2).

\subsection{Spectra in the ambient cloud outside cores}
\label{sec_mean_cloud}

To check if the two components, which are seen towards the coherent cores, are present throughout L1688, we looked at the spectra outside of the cores. Since the spectra outside of the cores are even fainter than those inside of them, we have to consider the average spectra in a small region, away from the cores. Therefore, we defined a rectangular box (shown by the black-dashed box in Figure \ref{coh_cores}) that is representative of the ambient cloud in L1688, as :
\begin{itemize}
    \item being at least two beams away from any coherent cores, to avoid contamination from the core spectra;
    \item not including, or being very close($\lesssim$ one beam) to known protostars; and
    \item being away from the western edge of the cloud, which is affected by a strong external illumination \citep{habart2003}.
\end{itemize}

\begin{figure*}[!ht]  
\centering
\includegraphics[width=0.8\textwidth]{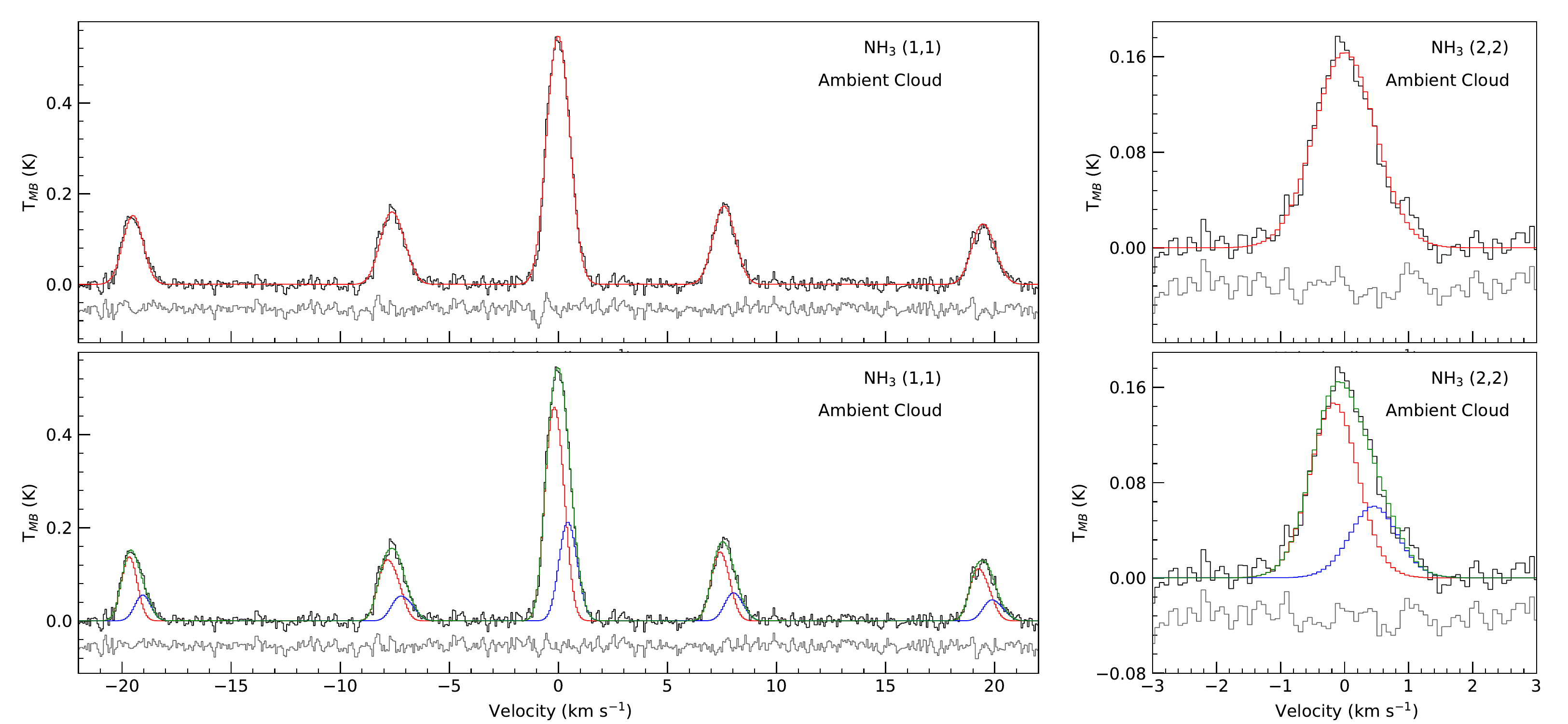}
  \caption{ {\it Top panels:} Average spectra of ambient cloud (defined in Section \ref{sec_mean_cloud}) with a single-component fit. {\it Bottom panels:} Same spectra, with two-component fit (green). The individual components are shown separately in red and blue.}
     \label{avg_spec_mean_cloud} 
\end{figure*}

The top panels of Figure \ref{avg_spec_mean_cloud} show the average spectrum in the ambient cloud with a single-component fit. Even when a two-component fit was attempted, we did not see a narrow component (bottom panel of Figure \ref{avg_spec_mean_cloud}). AIC indicates that the two-component fit is a better model, with both components being supersonic (Mach number >1). Therefore, we conclude that these are two broad components in the ambient cloud, which are separated in centroid velocity by $\approx$0.6 \kms.
The two components in the spectra have equal dispersions of 0.35 \kms. The kinetic temperatures of these two components are also approximately equal, 16.9 $\pm$ 0.2 K and 17.0 $\pm$ 0.5 K. These values are comparable to those from the broad components in the cores, and higher than the typical temperature of the narrow components. The Mach numbers of these two cloud components are comparable to that of the broad components seen in the cores (Table \ref{tab_fit_para}).  

The residuals from the single-component fit to the ambient cloud spectra are not as suggestive as in the cores. Therefore, the single-component model can also be considered as a reasonable fit. Even then, the comparison between the ambient cloud and the broad component towards the cores, as mentioned above, remains the same. The temperature, velocity dispersion, and the Mach number in the ambient cloud, from a single-component fit, are very similar to those in the broad component in the coherent cores.

\subsection{Comparison of physical parameters determined from the fit \label{sec_para_com}}

\begin{figure*}[!ht]  
\centering
\includegraphics[width=0.67\textwidth]{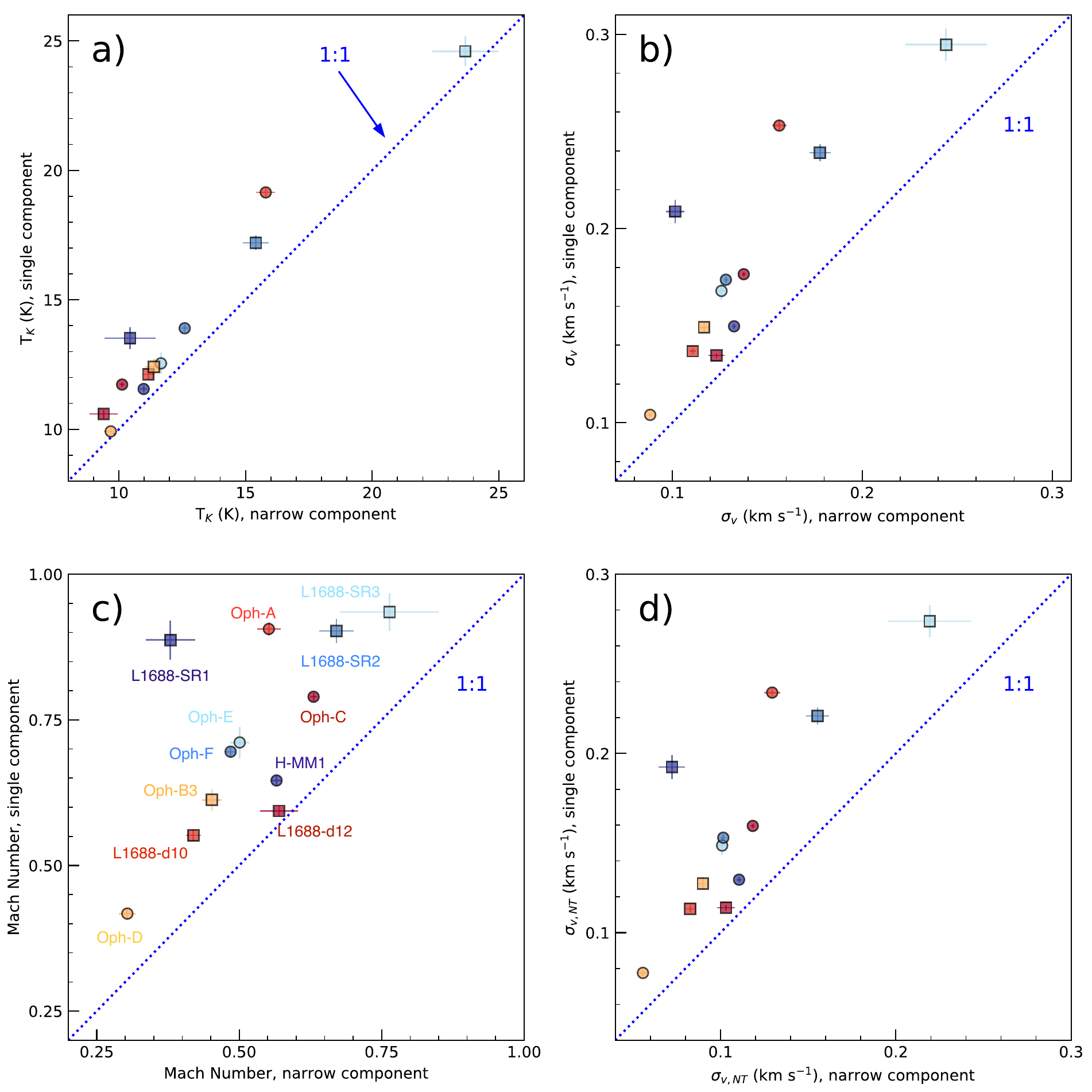}
\caption{Kinetic temperature, velocity dispersion, Mach number and non-thermal dispersion in the cores, from single-component fits, and for the narrow component of a two-component fit, are shown in panels a, b, c and d, respectively. 
In the cases where the error bar is not visible, the error is smaller than the symbols.}
 \label{para_2-1_cmp} 
\end{figure*}

Figure \ref{para_2-1_cmp} shows the comparison between kinetic temperature, velocity dispersion, Mach number, and non-thermal dispersion in the coherent cores, from a single-component fit, and for the narrow component in the two-component fit. Clearly, a single-component fit significantly overestimates all of these parameters in all cores. The kinetic temperature from the narrow component is lower by 1.19$\pm$0.07 K, compared to the single-component fit. Similarly, the velocity dispersion of the narrow component is lower by an average\footnote{Weighted averages of $\Delta$ \tk and $\Delta$\sig.} of 0.0332$\pm$ 0.0007 \kms. The difference is as high as $\approx$0.1 \kms in the case of SR1. On average, the single-component fit overestimates the kinetic temperature by 10\% and the velocity dispersion by 28\%, as compared to the narrow component. Correspondingly, the Mach number is overestimated, on average, by 32\%, and the non-thermal dispersion is typically overestimated by 39\%. 

The average kinetic temperatures and velocity dispersions inside the cores in the GAS DR1 parameter maps, which were obtained using single-component fits, are very similar to the single-component results we report. Therefore, the aforementioned comparison with narrow component results still holds, if the DR1 results (final two columns in Table \ref{tab_fit_para}) are considered. On average, the GAS DR1 results show a 14\% overestimate in \tk and a 34\% overestimate in \sig, as compared to our narrow component results.

\section{Discussion and conclusions}
\label{sec_diss}

For the first time, a faint, broad component has been detected towards coherent dense cores. When we consider a two-component fit to the spectra, it enables us to measure the velocity dispersion and kinetic temperature of the coherent core more accurately. We find that the typically-used single-component fit towards dense cores overestimates their temperatures and dispersions, on average, by 10\% and 28\%, respectively. 

The 1.2 K systematic offset in temperature in the cores derived from the single-component fit is important, as it is comparable to the temperature drop observed towards the centre of the cores when compared to the outer 1$\arcmin$ or 2$\arcmin$ \citep{crapsi_2007_temp_drop_core,harju_2017}. This decreased temperature of the cores has an effect on the chemistry inside the cores \citep{caselli_review_astrochem_2012, bergin_review_cold-cloud_2007}, since, at the volume densities of L1688 dense cores ($\sim10^5$ cm$^{-3}$), gas and dust are thermally coupled \citep{goldsmith_2001_dust_gas_coup} and surface chemistry rates exponentially depend on the dust temperature \citep[see][]{hasegawa_1992}.

Without the detection of the faint broad component, both the temperature and velocity dispersion in the cores are overestimated. This changes the estimates of the dynamical stability for the cores.

In the ambient cloud surrounding the coherent cores, we do not see the narrow component. Instead, at the position of our `ambient cloud box' (Figure \ref{coh_cores}), we see two broad components that are $\approx$0.6 \kms apart in centroid velocity, with equal temperatures and velocity dispersions. This might suggest the presence of two molecular clouds at slightly different velocities, which are probably merging at the location of L1688. The collision between these two clouds might result in a local density increase, where the narrow features are produced following a corresponding dissipation of turbulence, thus creating the observed coherent cores with subsonic linewidths.

On the other hand, if the faint broad features seen towards the coherent core position trace the less dense and more turbulent cloud along the line of sight of the coherent core, then we can measure the relative motions between the cores and the cloud in a direct way. 
Accordingly, we measured the relative velocity between the narrow and broad component, $\delta v_{(n-b)}$. Two-thirds of the coherent cores show subsonic $|\delta v_{(n-b)}|$ values, which suggests that most cores are dynamically coupled to their natal cloud \citep{kirk_2010_rel_vel}.
The standard deviation of $\delta v_{(n-b)}$ is 0.35 \kms, which is lower than the broad component's typical velocity dispersion (0.48 \kms). 
Therefore, the motions of the cores are lower than the typical motions in the surrounding gas \citep[see also][]{walsh_2007_rel_vel,kirk_2007_rel_vel,kirk_2010_rel_vel}. \citet{bailey_2015_rel_cor_vel} have also reported similar findings in simulated observations. 

The cloud velocity dispersions that we derived from \amm are lower than those obtained using lower density tracers (CO) in the earlier studies. Therefore, the results from lower density tracers might overestimate the local degree of turbulence in the cloud, or \amm based measurements provide a strict lower bound to it.

Without the required sensitivity in the observations, the faint broad component is not detected, and, therefore cannot be considered in the fit. This causes both the kinetic temperature and the velocity dispersion in the cores to be overestimated. Therefore, our results suggest that with deeper observations, we obtain better estimates of the core properties.

\begin{acknowledgements}
AP is supported by the Russian Ministry of Science and Higher Education via the State Assignment Project FEUZ-2020-0038. AP is a member of the Max Planck Partner Group at the Ural Federal University. SSRO acknowledges support from NSF CAREER grant. AC-T acknowledges support from MINECO project AYA2016-79006-P.
\end{acknowledgements}

\bibliographystyle{aa}
\bibliography{biblio}

\begin{thebibliography}{36}
\expandafter\ifx\csname natexlab\endcsname\relax\def\natexlab#1{#1}\fi

\bibitem[{{Bailey} {et~al.}(2015){Bailey}, {Basu}, \&
  {Caselli}}]{bailey_2015_rel_cor_vel}
{Bailey}, N.~D., {Basu}, S., \& {Caselli}, P. 2015, \apj, 798, 75

\bibitem[{{Barranco} \& {Goodman}(1998)}]{coh_core_barranco_goodman_1998}
{Barranco}, J.~A. \& {Goodman}, A.~A. 1998, \apj, 504, 207

\bibitem[{{Bergin} \& {Tafalla}(2007)}]{bergin_review_cold-cloud_2007}
{Bergin}, E.~A. \& {Tafalla}, M. 2007, \araa, 45, 339

\bibitem[{{Caselli} {et~al.}(2002){Caselli}, {Benson}, {Myers}, \&
  {Tafalla}}]{caselli_2002_dense_core}
{Caselli}, P., {Benson}, P.~J., {Myers}, P.~C., \& {Tafalla}, M. 2002, \apj,
  572, 238

\bibitem[{{Caselli} {et~al.}(2017){Caselli}, {Bizzocchi}, {Keto}, {Sipil{\"a}},
  {Tafalla}, {Pagani}, {Kristensen}, {van der Tak}, {Walmsley}, {Codella},
  {Nisini}, {Aikawa}, {Faure}, \& {van Dishoeck}}]{caselli_2017_amm}
{Caselli}, P., {Bizzocchi}, L., {Keto}, E., {et~al.} 2017, \aap, 603, L1

\bibitem[{{Caselli} \& {Ceccarelli}(2012)}]{caselli_review_astrochem_2012}
{Caselli}, P. \& {Ceccarelli}, C. 2012, \aapr, 20, 56

\bibitem[{{Chen} {et~al.}(2019){Chen}, {Pineda}, {Goodman}, {Burkert},
  {Offner}, {Friesen}, {Myers}, {Alves}, {Arce}, \& {Caselli}}]{chen2019}
{Chen}, H. H.-H., {Pineda}, J.~E., {Goodman}, A.~A., {et~al.} 2019, \apj, 877,
  93

\bibitem[{{Crapsi} {et~al.}(2007){Crapsi}, {Caselli}, {Walmsley}, \&
  {Tafalla}}]{crapsi_2007_temp_drop_core}
{Crapsi}, A., {Caselli}, P., {Walmsley}, M.~C., \& {Tafalla}, M. 2007, \aap,
  470, 221

\bibitem[{{Di Francesco} {et~al.}(2004){Di Francesco}, {Andr{\'e}}, \&
  {Myers}}]{di_fran_2004_ophA_n}
{Di Francesco}, J., {Andr{\'e}}, P., \& {Myers}, P.~C. 2004, \apj, 617, 425

\bibitem[{{Dunham} {et~al.}(2015){Dunham}, {Allen}, {Evans},
  {Broekhoven-Fiene}, {Cieza}, {Di Francesco}, {Gutermuth}, {Harvey},
  {Hatchell}, {Heiderman}, {Huard}, {Johnstone}, {Kirk}, {Matthews}, {Miller},
  {Peterson}, \& {Young}}]{yso_l1688}
{Dunham}, M.~M., {Allen}, L.~E., {Evans}, Neal~J., I., {et~al.} 2015, \apjs,
  220, 11

\bibitem[{{Friesen} {et~al.}(2009){Friesen}, {Di Francesco}, {Shirley}, \&
  {Myers}}]{friesen_2009_b3}
{Friesen}, R.~K., {Di Francesco}, J., {Shirley}, Y.~L., \& {Myers}, P.~C. 2009,
  \apj, 697, 1457

\bibitem[{{Friesen} {et~al.}(2017){Friesen}, {Pineda}, {co-PIs}, {Rosolowsky},
  {Alves}, {Chac{\'o}n-Tanarro}, {How-Huan Chen}, {Chun-Yuan Chen}, {Di
  Francesco}, {Keown}, {Kirk}, {Punanova}, {Seo}, {Shirley}, {Ginsburg},
  {Hall}, {Offner}, {Singh}, {Arce}, {Caselli}, {Goodman}, {Martin}, {Matzner},
  {Myers}, {Redaelli}, \& {GAS Collaboration}}]{GASDR1}
{Friesen}, R.~K., {Pineda}, J.~E., {co-PIs}, {et~al.} 2017, \apj, 843, 63

\bibitem[{{Ginsburg} \& {Mirocha}(2011)}]{pyspeckit}
{Ginsburg}, A. \& {Mirocha}, J. 2011, {PySpecKit: Python Spectroscopic Toolkit}

\bibitem[{{Goldsmith}(2001)}]{goldsmith_2001_dust_gas_coup}
{Goldsmith}, P.~F. 2001, \apj, 557, 736

\bibitem[{{Goodman} {et~al.}(1998){Goodman}, {Barranco}, {Wilner}, \&
  {Heyer}}]{coh_core_goodman_barranco_1998}
{Goodman}, A.~A., {Barranco}, J.~A., {Wilner}, D.~J., \& {Heyer}, M.~H. 1998,
  \apj, 504, 223

\bibitem[{{Habart} {et~al.}(2003){Habart}, {Boulanger}, {Verstraete}, {Pineau
  des For{\^e}ts}, {Falgarone}, \& {Abergel}}]{habart2003}
{Habart}, E., {Boulanger}, F., {Verstraete}, L., {et~al.} 2003, \aap, 397, 623

\bibitem[{{Harju} {et~al.}(2017){Harju}, {Daniel}, {Sipil{\"a}}, {Caselli},
  {Pineda}, {Friesen}, {Punanova}, {G{\"u}sten}, {Wiesenfeld}, {Myers},
  {Faure}, {Hily-Blant}, {Rist}, {Rosolowsky}, {Schlemmer}, \&
  {Shirley}}]{harju_2017}
{Harju}, J., {Daniel}, F., {Sipil{\"a}}, O., {et~al.} 2017, \aap, 600, A61

\bibitem[{{Hasegawa} {et~al.}(1992){Hasegawa}, {Herbst}, \&
  {Leung}}]{hasegawa_1992}
{Hasegawa}, T.~I., {Herbst}, E., \& {Leung}, C.~M. 1992, \apjs, 82, 167

\bibitem[{{Johnstone} {et~al.}(2004){Johnstone}, {Di Francesco}, \&
  {Kirk}}]{johnstone_2004_hmm1}
{Johnstone}, D., {Di Francesco}, J., \& {Kirk}, H. 2004, \apjl, 611, L45

\bibitem[{{Kauffmann} {et~al.}(2008){Kauffmann}, {Bertoldi}, {Bourke}, {Evans},
  \& {Lee}}]{jens_2008_mu}
{Kauffmann}, J., {Bertoldi}, F., {Bourke}, T.~L., {Evans}, N.~J., I., \& {Lee},
  C.~W. 2008, \aap, 487, 993

\bibitem[{{Kirk} {et~al.}(2007){Kirk}, {Johnstone}, \&
  {Tafalla}}]{kirk_2007_rel_vel}
{Kirk}, H., {Johnstone}, D., \& {Tafalla}, M. 2007, \apj, 668, 1042

\bibitem[{{Kirk} {et~al.}(2010){Kirk}, {Pineda}, {Johnstone}, \&
  {Goodman}}]{kirk_2010_rel_vel}
{Kirk}, H., {Pineda}, J.~E., {Johnstone}, D., \& {Goodman}, A. 2010, \apj, 723,
  457

\bibitem[{{Koch} {et~al.}(2018){Koch}, {Rosolowsky}, \&
  {Leroy}}]{koch_2018_chan-width}
{Koch}, E., {Rosolowsky}, E., \& {Leroy}, A.~K. 2018, Research Notes of the
  American Astronomical Society, 2, 220

\bibitem[{{Ladjelate} {et~al.}(2020){Ladjelate}, {Andr{\'e}}, {K{\"o}nyves},
  {Ward-Thompson}, {Men'shchikov}, {Bracco}, {Palmeirim}, {Roy}, {Shimajiri},
  {Kirk}, {Arzoumanian}, {Benedettini}, {Di Francesco}, {Fiorellino},
  {Schneider}, \& {Pezzuto}}]{ladj_2020_hgbs_oph_strc}
{Ladjelate}, B., {Andr{\'e}}, P., {K{\"o}nyves}, V., {et~al.} 2020, arXiv
  e-prints, arXiv:2001.11036

\bibitem[{{Leroy} {et~al.}(2016){Leroy}, {Hughes}, {Schruba}, {Rosolowsky},
  {Blanc}, {Bolatto}, {Colombo}, {Escala}, {Kramer}, {Kruijssen}, {Meidt},
  {Pety}, {Querejeta}, {Sandstrom}, {Schinnerer}, {Sliwa}, \&
  {Usero}}]{leroy_2016_chan-width}
{Leroy}, A.~K., {Hughes}, A., {Schruba}, A., {et~al.} 2016, \apj, 831, 16

\bibitem[{{Loren} {et~al.}(1990){Loren}, {Wootten}, \&
  {Wilking}}]{loren_1990_b3}
{Loren}, R.~B., {Wootten}, A., \& {Wilking}, B.~A. 1990, \apj, 365, 269

\bibitem[{{Mangum} {et~al.}(2007){Mangum}, {Emerson}, \&
  {Greisen}}]{mangum_2007_kernel}
{Mangum}, J.~G., {Emerson}, D.~T., \& {Greisen}, E.~W. 2007, \aap, 474, 679

\bibitem[{{Markwardt}(2009)}]{markwardt2009}
{Markwardt}, C.~B. 2009, in Astronomical Society of the Pacific Conference
  Series, Vol. 411, Astronomical Data Analysis Software and Systems XVIII, ed.
  D.~A. {Bohlender}, D.~{Durand}, \& P.~{Dowler}, 251

\bibitem[{{Motte} {et~al.}(1998){Motte}, {Andre}, \& {Neri}}]{motte1998}
{Motte}, F., {Andre}, P., \& {Neri}, R. 1998, \aap, 336, 150

\bibitem[{{Myers}(1983)}]{myers_1983_SF_core}
{Myers}, P.~C. 1983, \apj, 270, 105

\bibitem[{{Myers} \& {Benson}(1983)}]{myers-benson_1983_dense_core}
{Myers}, P.~C. \& {Benson}, P.~J. 1983, \apj, 266, 309

\bibitem[{{Ortiz-Le{\'o}n} {et~al.}(2018){Ortiz-Le{\'o}n}, {Loinard}, {Dzib},
  {Kounkel}, {Galli}, {Tobin}, {Evans}, {Hartmann}, {Rodr{\'\i}guez},
  {Brice{\~n}o}, {Torres}, \& {Mioduszewski}}]{l1688_dist_ortiz-leon}
{Ortiz-Le{\'o}n}, G.~N., {Loinard}, L., {Dzib}, S.~A., {et~al.} 2018, \apjl,
  869, L33

\bibitem[{{Pineda} {et~al.}(2010){Pineda}, {Goodman}, {Arce}, {Caselli},
  {Foster}, {Myers}, \& {Rosolowsky}}]{pineda2010}
{Pineda}, J.~E., {Goodman}, A.~A., {Arce}, H.~G., {et~al.} 2010, \apj, 712,
  L116

\bibitem[{{Rosolowsky} {et~al.}(2008){Rosolowsky}, {Pineda}, {Foster},
  {Borkin}, {Kauffmann}, {Caselli}, {Myers}, \&
  {Goodman}}]{roso_2008_amm_dens_core}
{Rosolowsky}, E.~W., {Pineda}, J.~E., {Foster}, J.~B., {et~al.} 2008, \apjs,
  175, 509

\bibitem[{{Walsh} {et~al.}(2007){Walsh}, {Myers}, {Di Francesco}, {Mohanty},
  {Bourke}, {Gutermuth}, \& {Wilner}}]{walsh_2007_rel_vel}
{Walsh}, A.~J., {Myers}, P.~C., {Di Francesco}, J., {et~al.} 2007, \apj, 655,
  958

\bibitem[{{Williams} \& {Myers}(2000)}]{williams_2000_coher}
{Williams}, J.~P. \& {Myers}, P.~C. 2000, \apj, 537, 891

\end{thebibliography}

\appendix

\section{Calculation of non-thermal velocity dispersion, $\sigma_{\rm NT}$}
\label{app_sig_nt_calc}

The non-thermal component of the velocity dispersion is calculated by removing the thermal dispersion for the observed molecule ($\rm{\sigma_T}$) from the total observed velocity dispersion (\sig) :
\begin{equation}
    {\sigma_{NT}}^2 = {\sig}^2 - {\sigma_{T,\amm}}^2 -{\sigma_{chan}}^2~,
\end{equation}
where $\sigma_{chan}$ is the contribution due to the width of the channel 
and the thermal component of the dispersion observed in \amm is 
\begin{equation}
    \sigma_{T,\amm} = \sqrt{\frac{k_B T_k}{\mu_{\amm}}}~,
\end{equation}
with $\rm{\mu_{\amm}}$ = 17 amu, being the mass of the ammonia molecule. 
The typical values of \tk and \sig in this study are 14\,K and 0.2\,\kms, respectively (see Section~\ref{sec_comp_1-2_comp}). At \tk=14\,K, the thermal component, $\sigma_{T,\amm}$ is 0.082 \kms.

When a Hann-like kernel is used in the spectrometer, then the correction factor for the measured velocity dispersion is given by \citep[see][]{leroy_2016_chan-width, koch_2018_chan-width} 
\begin{equation}
    \sigma_{chan} = \frac{\Delta}{\sqrt{2 \pi}} (1.0 + 1.18k + 10.4 k^2)~,
\end{equation}
where $\Delta$ is the spectral resolution and $k$ is dependent on the Hann-like function applied.
In the case of the VEGAS spectrometer used in the GAS observations, k is 0.11, which corresponds to a correction term of
\begin{equation}
    \sigma_{chan} = \frac{\Delta}{1.994} = 0.036\, \kms ~.
\end{equation}
After applying this small correction, we obtain a typical non-thermal component of the velocity dispersion, $\sigma_{NT}=0.182$ \kms.

\section{{Coherent cores}}
\label{app_core_names}

{As mentioned in Section \ref{sec_id_coh}, we identify 12 regions in L1688 as coherent cores. These include}:
\begin{itemize}
    \item five cores, Oph-A, Oph-C, Oph-D, Oph-E and Oph-F, which were identified in continuum emission \citep{motte1998},
    \item one DCO$^+$ core, Oph-B3 \citep{loren_1990_b3, friesen_2009_b3}
    \item two coherent `droplets', L1688-d10 and L1688-d12, \citep{chen2019}, and
    \item one prestellar core, Oph-H-MM1 \citep{johnstone_2004_hmm1}.
\end{itemize} 

{Apart from these previously identified cores, we also identified }three more subsonic regions :  L1688-SR1 \citep[south of Oph-B1 in][]{motte1998}, L1688-SR2 (three islands east of Oph-A), and L1688-SR3 (near the western edge of the map). {\citet{ladj_2020_hgbs_oph_strc} identifies a prestellar core at the position of SR1 and an unbound starless core at the position of SR3. SR1 and SR3 are associated with a local peak in \amm integrated intensity. SR2 does not contain any bound core in the Herschel data, nor does it show a local \amm peak.}

Oph-B3, Oph-D and H-MM1 are also identified as L1688-c4, L1688-d11 and L1688-d9, respectively, in \citet{chen2019}. The two structures that we identify as Oph-A are likely associated with Oph A-N2 and Oph A-N6, as identified in N$_2$H$^+$ {by} \citet{di_fran_2004_ophA_n}. 
The continuum cores Oph-B1 and Oph-B2 from \citet{motte1998} are not subsonic {\citep{friesen_2009_b3} } and, therefore, they are not considered as `coherent cores' by our definition.

\section{Checking effect of smoothing the data on the final results}
\label{app_chk_unsmth}

To check if there is any effect due to smoothing the data (see Section \ref{sec_amm_data}), we performed the same averaging on the original GAS data, using the velocity map published in DR1 for aligning the spectra. Since the DR1 velocity map was not as extensive, this test could not be performed for all the cores, as we need the velocity information to align the spectra. In the cores where this check was possible, the results from the new fits were  not significantly different ($\approx$1.4\% change in \tk and $\approx$4.7\% change in \sig). In comparison, the difference in the estimates from the narrow component and from single-component fit results were 10\% in \tk and 28\% in \sig (Section \ref{sec_para_com}).

\section{Fitting the average spectra in the cores}
\label{app_avg_fit}

To fit the average spectra in cores, we followed the same process as described in Section \ref{sec_line_fit}. We used the module \texttt{specfit} in \texttt{pyspeckit} to fit single spectra. Since we were fitting the spectra towards the cores, we used, as our initial guess, slightly lower values of 12\,K and 4\,K for \tk and \tex, respectively. As the spectra are aligned at v=0, we used \vel=0.0 \kms as our initial guess. For \sig, we used the mean value inside all cores, which is \sig $\approx$ 0.19 \kms, from the extended velocity dispersion map obtained from the fit process described in Section \ref{sec_line_fit}.

While fitting two components to the spectra, we added a two-component \texttt{cold\_ammonia} model as a new model in the fitter. We then used this model to fit the spectra. As the residuals from single-component fits suggest the presence of a narrow and a broad component, we used two values, one from each side of the guess used in the single-component fit (0.19 \kms) as our initial \sig guesses for the two components. After checking the fit with varying initial guesses, we found that small differences in \tk and \vel in the initial guesses also aid in arriving at a stable fit. Therefore, we used the following initial guesses in the two-component fits: 
\begin{multicols}{2}
\begin{itemize}
    \item (\tk)$_1$ = 11 K
    \item (\tex)$_1$ = 4 K
    \item (\sig)$_1$ = 0.15 \kms
    \item (\vel)$_1$ = 0.0 \kms
    \item (\tk)$_2$ = 16 K
    \item (\tex)$_2$ = 4 K
    \item (\sig)$_2$ = 0.3 \kms
    \item (\vel)$_2$ = 0.25 \kms
\end{itemize}
\end{multicols}

\section{Model selection using AIC estimator 
\label{app_mod_sel}}

We fitted the averaged spectra considering models with a different number of components in ammonia, with each component being modelled as a \texttt{cold\_ammonia} spectrum in \texttt{pyspeckit}. The number of parameters used in the models is five per component. To select the best model to fit a spectra, we checked the Akaike Information Criterion (AIC), which determines if the quality of the model significantly improves considering the increase in the number of parameters used. 
Assuming that each channel in the spectra has a constant Gaussian error, which is equal to the noise in the spectra, AIC is related to the $\chi^2$ of the data as :
\begin{equation}\label{eq_chi_aic}
    AIC = 2k + \chi^2 + C,
\end{equation}
where $k$ is the number of parameters used to model the data and C is a constant, which depends on the noise in the spectra, and is given by :
\begin{equation}
C = - 2 N\, \times\, 
\ln\left(\frac{1}{\sqrt{2 \pi \sigma^2}}\right),
\end{equation}
where N is the number of channels in the spectra.

The model with the lowest AIC value is considered to be the best model. In our case, the channels are not completely independent and, therefore the actual number of independent channels is different. This effect, however, is systematic and it does not change for models with different number of components. The change in the AIC value from one model to another is the important term, and the actual AIC values are less significant.

Using Equation \ref{eq_chi_aic} we can rewrite the change in AIC value from a single-component fit to a two-component fit, $\Delta_{\rm AIC}$, as
\begin{equation}
    \Delta_{\rm AIC} = 2 \Delta k + \Delta \chi^2 ~,
\end{equation}
where $\Delta k$ is the change in number of parameters and $\Delta \chi^2$ is the change in $\chi^2$, from a single-component fit to a two-component fit.
Each model component uses five independent parameters and, therefore $\rm \Delta k = -5$. 
Also, $\rm \Delta \chi^2$ can be divided into two groups: channels with and without emission, or 
\begin{equation}
    \Delta \chi^2 = (\Delta \chi^2)_{with\ emission} + (\Delta \chi^2)_{without\ emission}~.
\end{equation}
Outside the emission region the models are identical and, therefore, 
\begin{equation}
    \Delta \chi^2 
    = \frac{n_{line}}{\sigma^2} 
    \left(rms_{emission,\ 1-comp}^2 - 
    rms_{emission,\ 2-comp}^2\right)~,
\end{equation}
where, $\rm n_{line}$ is the number of channels with emission, $\sigma$ is the noise in the spectra, and $rms_{\rm emission}$ is the residual root mean square (rms) of the model in the emission region.
The comparison of the $\sigma_{\rm emission}$ and the noise, $\sigma$, is taken as a measure of determining the goodness of the fit in line-fitting software, such as CLASS in GILDAS\footnote{\url{http://www.iram.fr/IRAMFR/GILDAS}}.
Then, in our comparison of single- and two-component fits, we have 
\begin{equation}
\label{eq_rms_aic}
    \Delta_{AIC} = -10 + \frac{n_{line}}{\sigma^2} \left(rms_{emission,\ 1-comp}^2 - 
    rms_{emission,\ 2-comp}^2\right)~.
\end{equation}

We note that$\rm n_{line}$ depends on the width of the spectra and, therefore, it is different for each coherent core. From Equation \ref{eq_rms_aic}, we see that for a low variation in the AIC, the improvement in the residuals of the fit is also small. We find that for the two coherent cores with low $\rm \Delta_{AIC}$ (<30), d12 and SR3, the improvement in the $\rm rms_{emission}$ is less than 10\% of the noise in the respective spectra. For the ambient cloud spectra, with $\rm \Delta_{AIC} = $94, $\rm rms_{emission}$ improves by 10\% of the noise level, from a single-component to a two-component fit. For every other region, we see an improvement in $\rm rms_{emission}$ by >15\%. 

Owing to our limited number of data points, it is not possible to definitively determine what change in AIC represents a significant improvement in the fit upon the addition of the second component. From the corresponding improvement in $\rm rms_{emission}$, we note that in our current work, only the regions with $\rm \Delta_{AIC} > 150$ show a significant improvement in the fit.

\section{Average spectra of the whole cloud}
\label{app_avg_spec_whole}

\begin{figure*}[!ht]  
\centering
\includegraphics[width=0.98\textwidth]{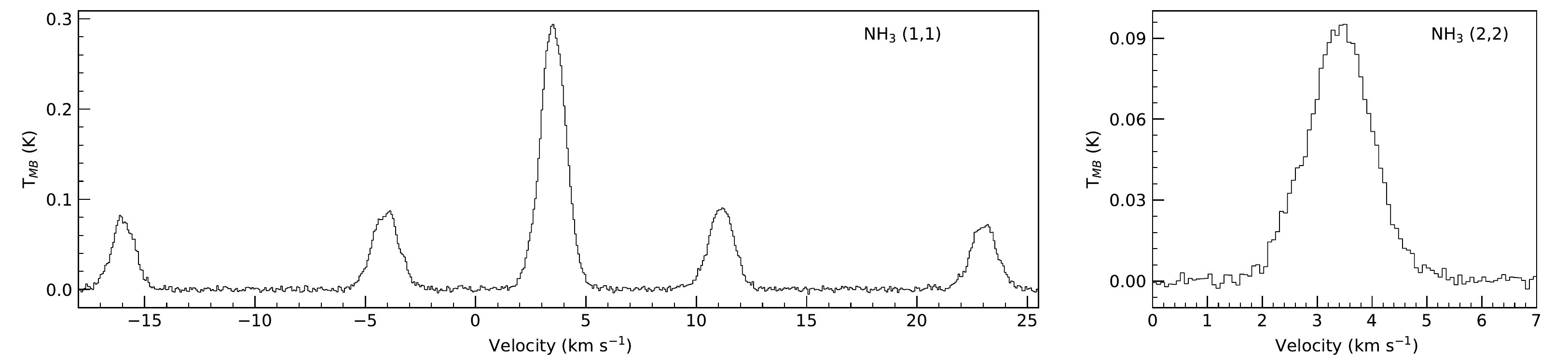}
  \caption{Average \amm (1,1) and (2,2) spectra in L1688.}
     \label{avg_spec_whole} 
\end{figure*}

Figure \ref{avg_spec_whole} shows the average (1,1) and (2,2) spectra in the whole L1688, which was mapped by the Green Bank Ammonia Survey (GAS). These spectra were used to determine the range of centroid velocity, for which a fit is attempted with \verb+pyspeckit+.

\section{Averaged spectra towards cores}
\label{app_avg_spec}

\begin{figure*}[!ht]  
\centering
\includegraphics[width=0.98\textwidth]{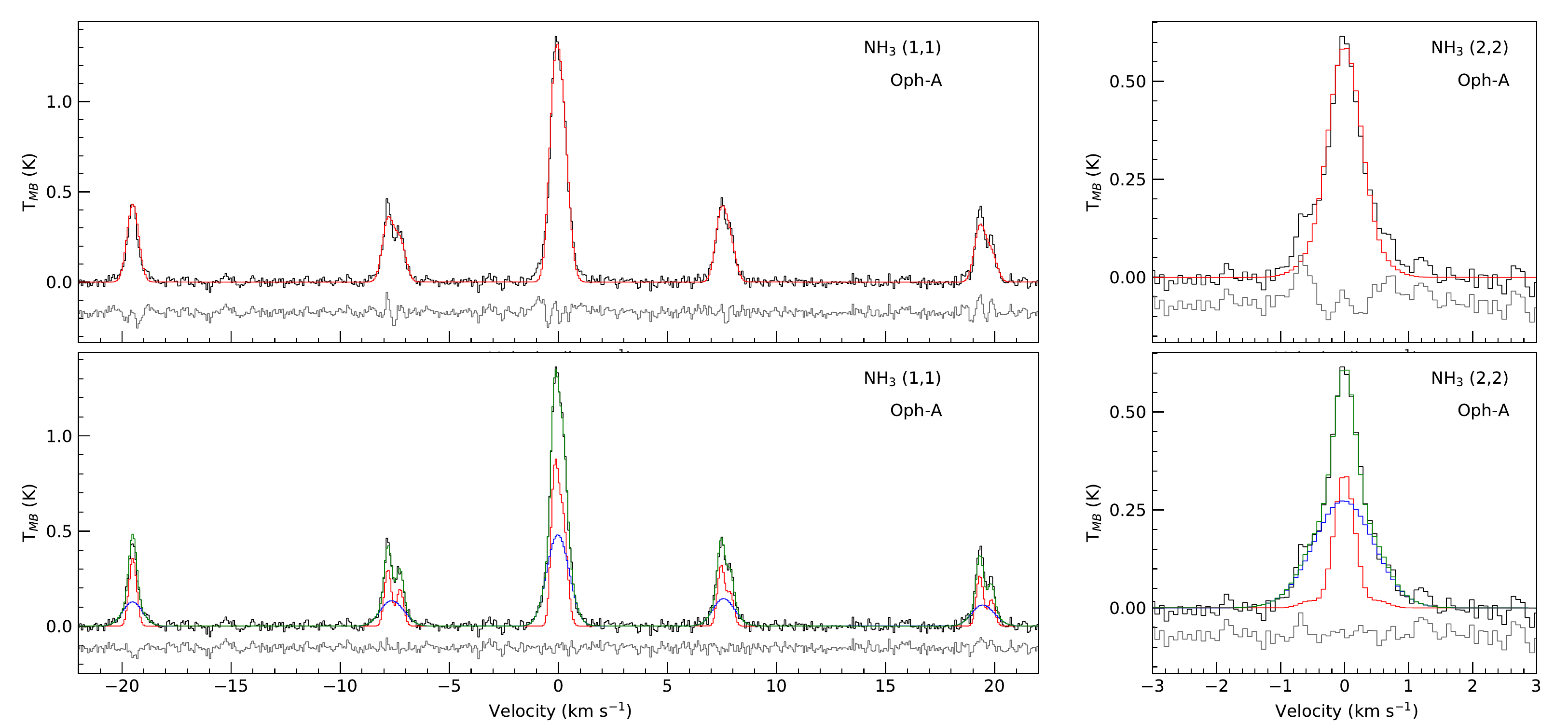}
  \caption{Top panels: Average \amm (1,1) and (2,2) spectra of Oph-A core, with a single-component fit. Bottom panels: Same spectra, with two-component fit (green). The narrow (red) and broad (blue) components are also shown separately.}
     \label{avg_spec_ophA} 
\end{figure*}

\begin{figure*}[!ht]  
\centering
\includegraphics[width=0.98\textwidth]{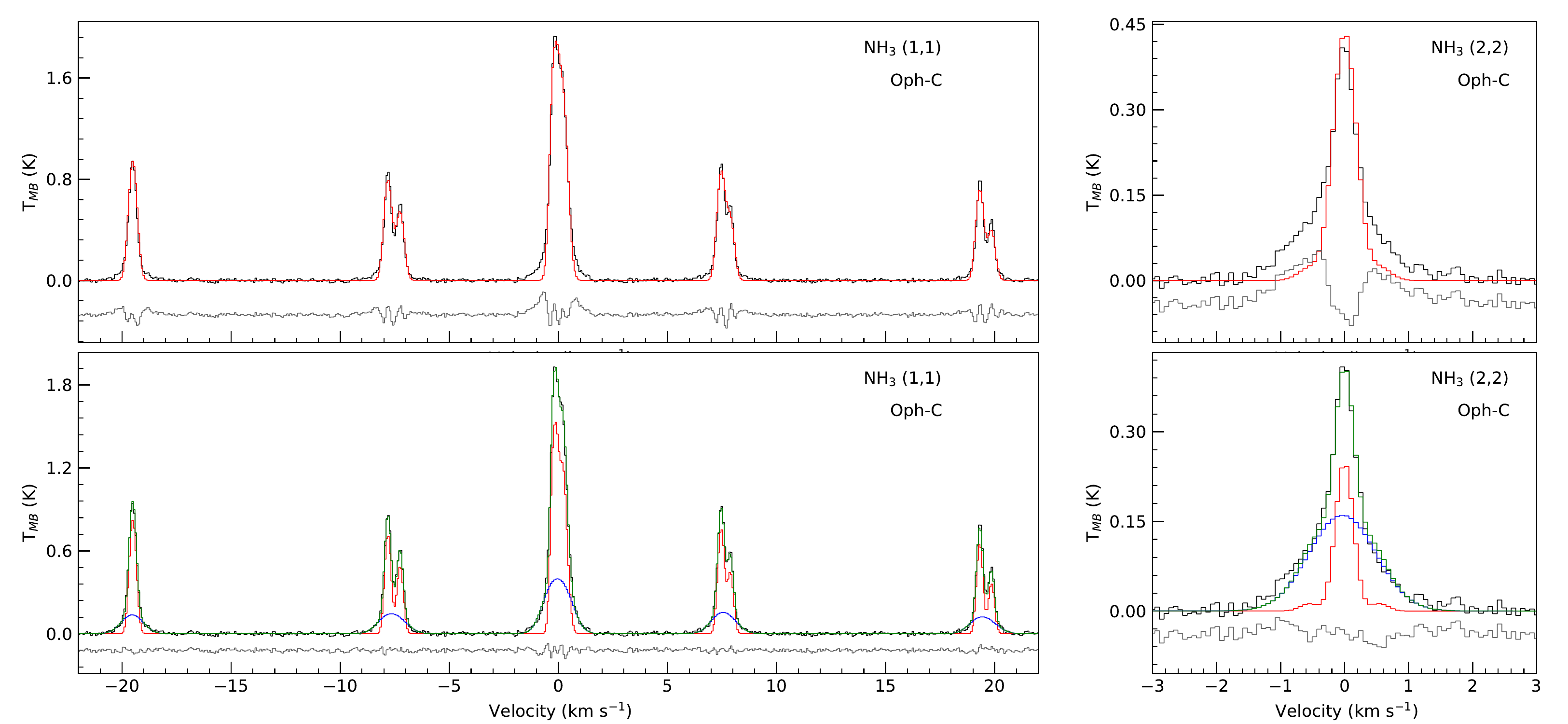}
  \caption{Same as Figure \ref{avg_spec_ophA}, but for Oph-C.}
     \label{avg_spec_ophC} 
\end{figure*}

\begin{figure*}[!ht]  
\centering
\includegraphics[width=0.98\textwidth]{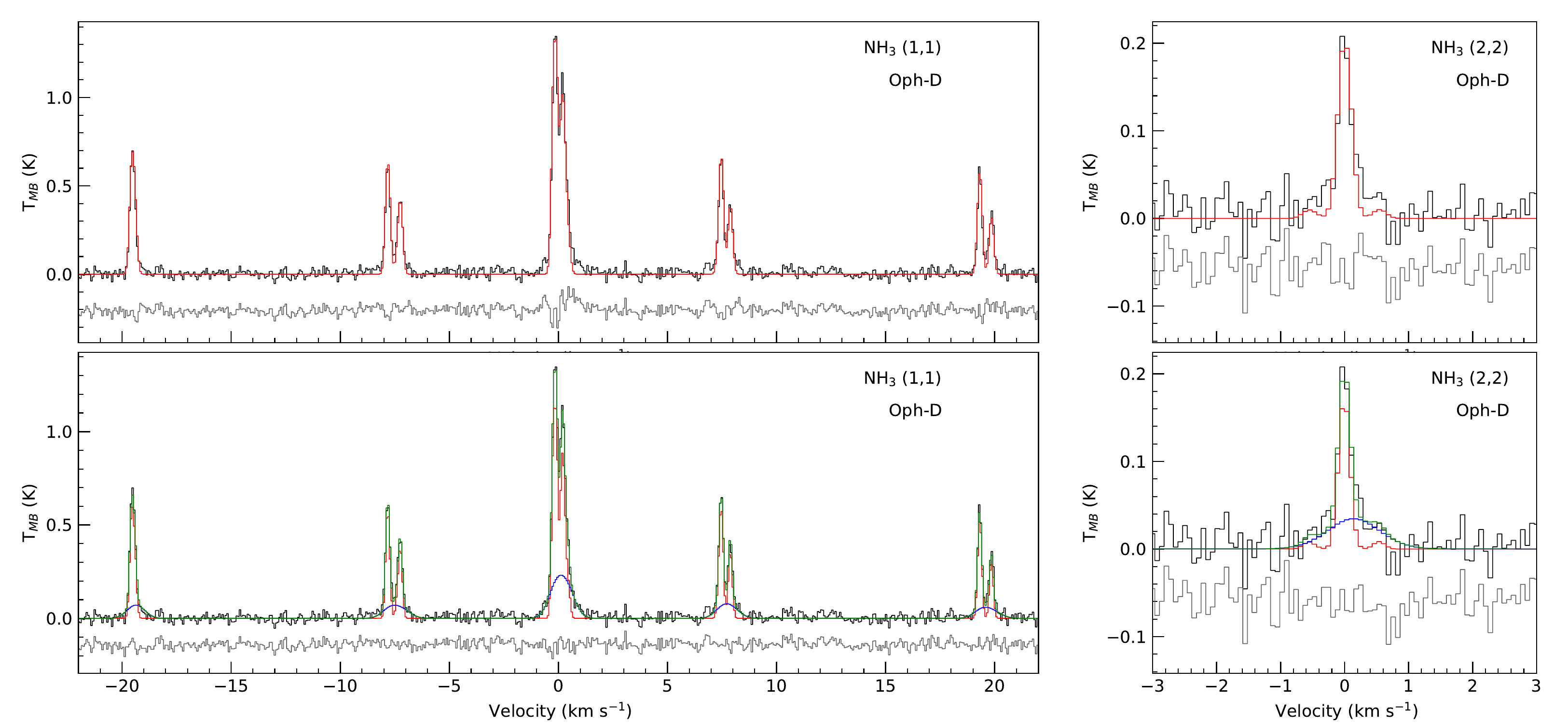}
  \caption{Same as Figure \ref{avg_spec_ophA}, but for Oph-D.}
     \label{avg_spec_ophD} 
\end{figure*}

\begin{figure*}[!ht]  
\centering
\includegraphics[width=0.98\textwidth]{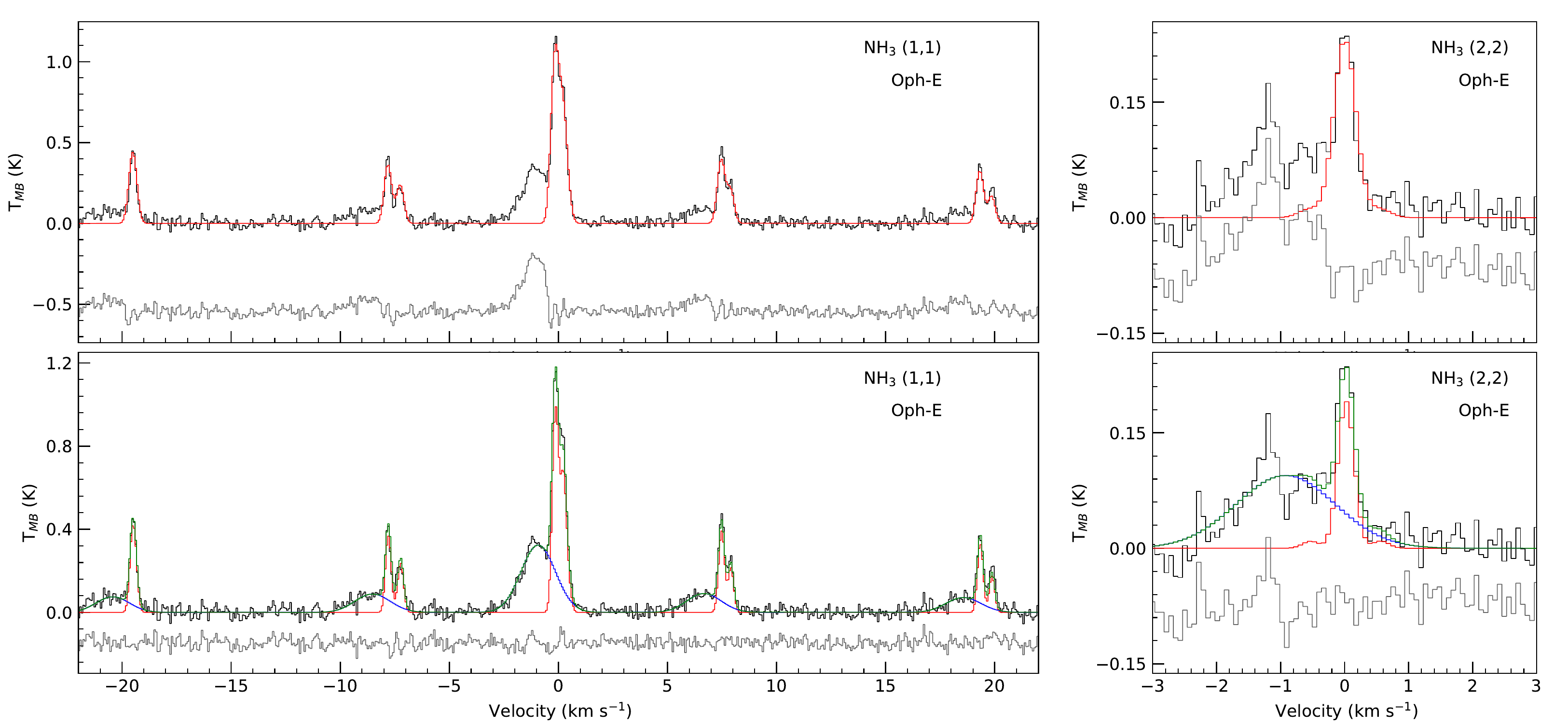}
  \caption{Same as Figure \ref{avg_spec_ophA}, but for Oph-E.}
     \label{avg_spec_ophE} 
\end{figure*}

\begin{figure*}[!ht]  
\centering
\includegraphics[width=0.98\textwidth]{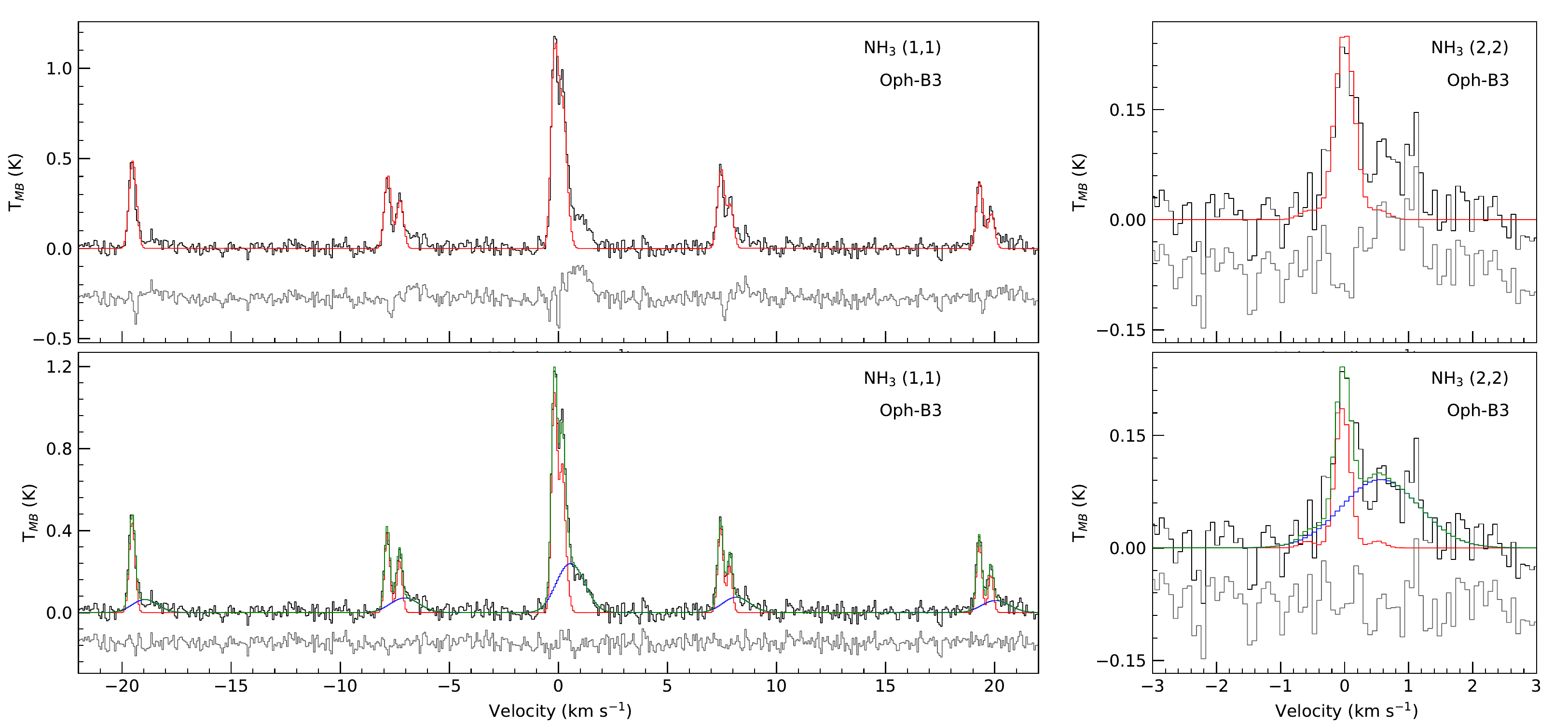}
  \caption{Same as Figure \ref{avg_spec_ophA}, but for Oph-B3.}
     \label{avg_spec_c4} 
\end{figure*}

\begin{figure*}[!ht]  
\centering
\includegraphics[width=0.98\textwidth]{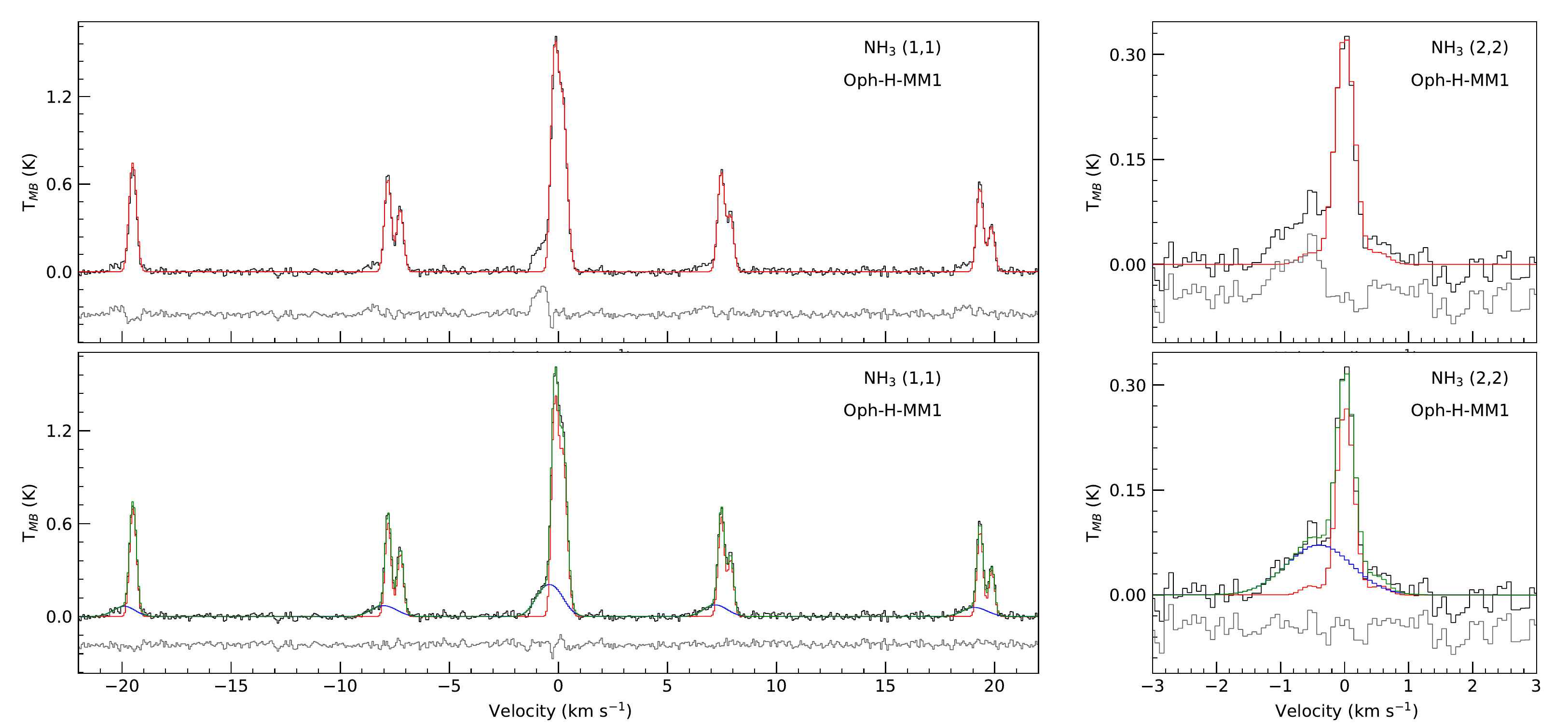}
  \caption{Same as Figure \ref{avg_spec_ophA}, but for Oph-H-MM1.}
     \label{avg_spec_hmm1} 
\end{figure*}

\begin{figure*}[!ht]  
\centering
\includegraphics[width=0.98\textwidth]{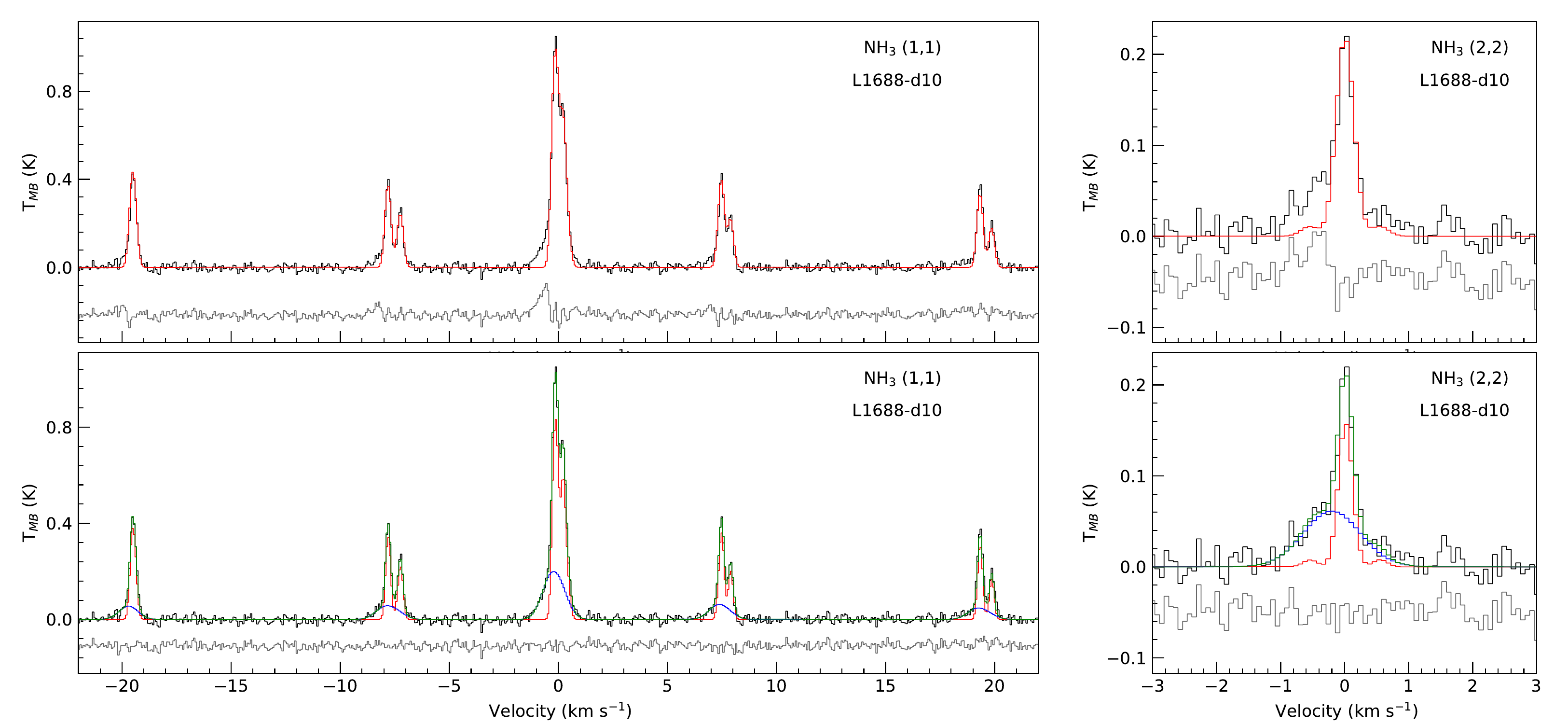}
  \caption{Same as Figure \ref{avg_spec_ophA}, but for L1688-d10.}
     \label{avg_spec_d10} 
\end{figure*}

\begin{figure*}[!ht]  
\centering
\includegraphics[width=0.98\textwidth]{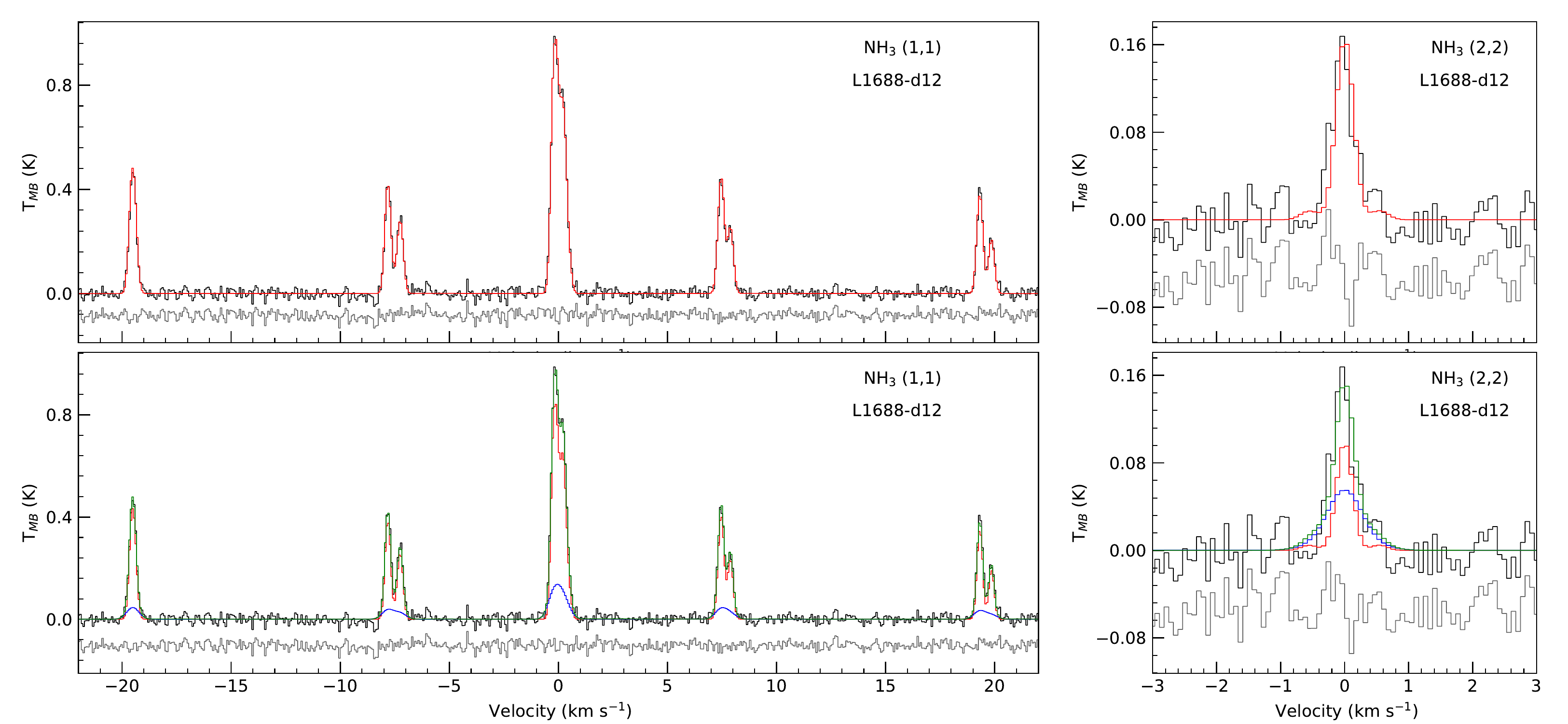}
  \caption{Same as Figure \ref{avg_spec_ophA}, but for L1688-d12.}
     \label{avg_spec_d12} 
\end{figure*}

\begin{figure*}[!ht]  
\centering
\includegraphics[width=0.98\textwidth]{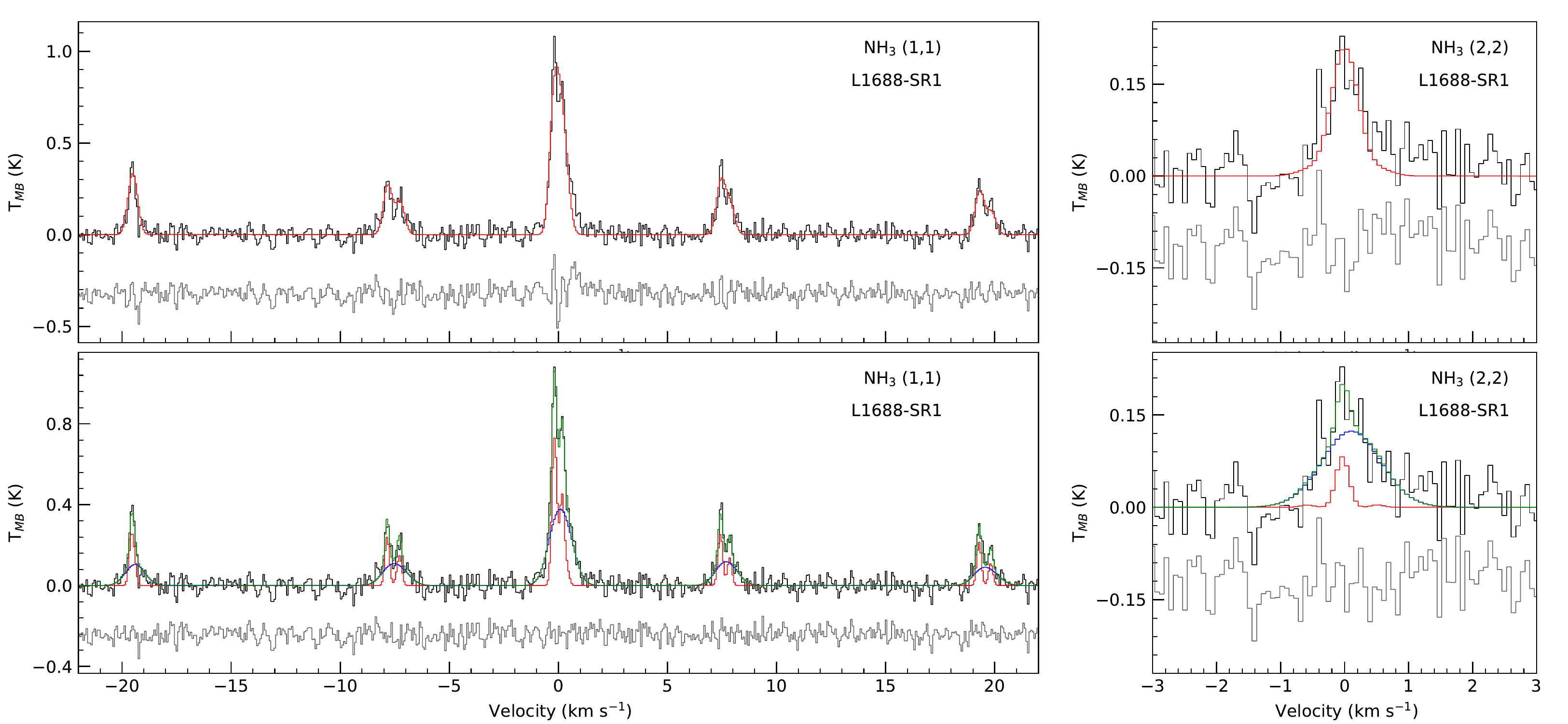}
  \caption{Same as Figure \ref{avg_spec_ophA}, but for L1688-SR1.}
     \label{avg_spec_bsou} 
\end{figure*}

\begin{figure*}[!ht]  
\centering
\includegraphics[width=0.98\textwidth]{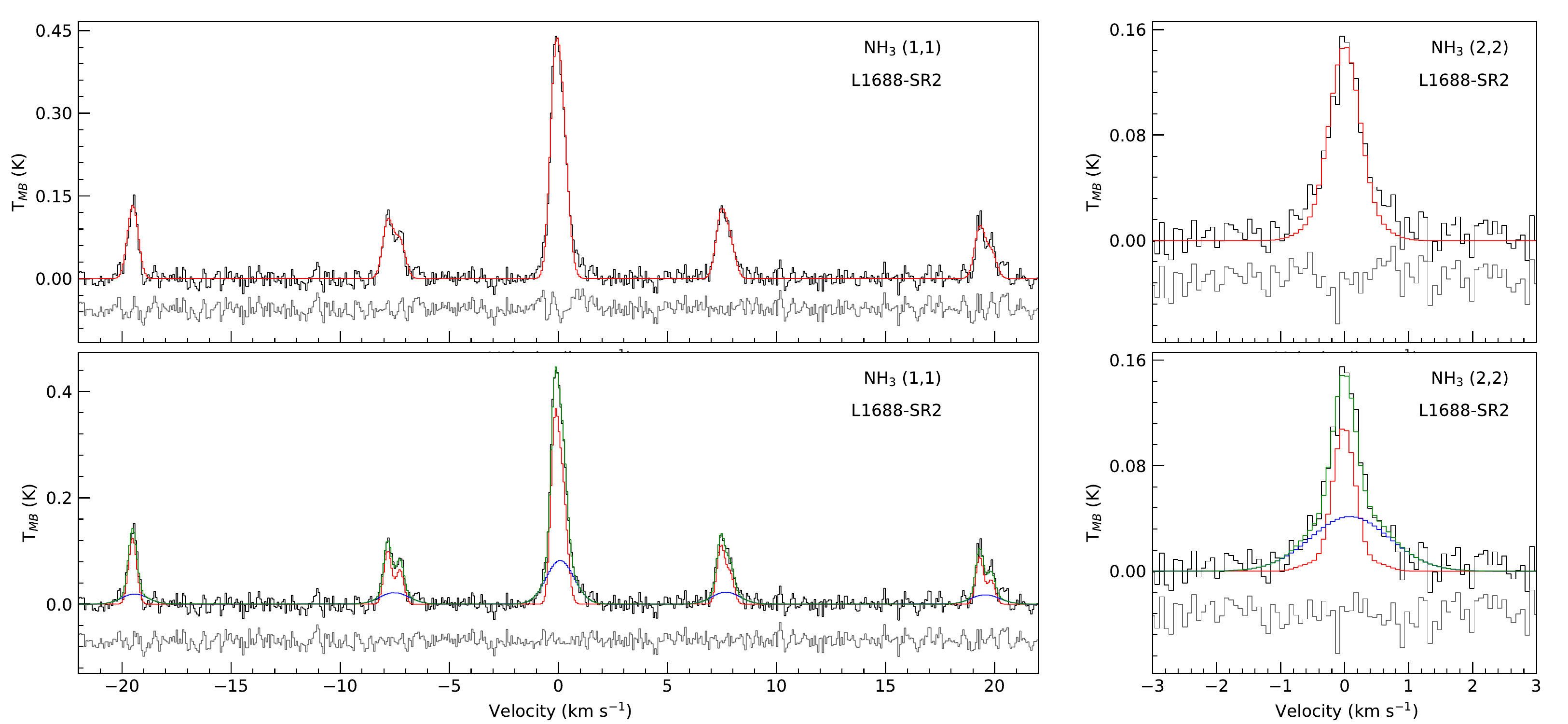}
  \caption{Same as Figure \ref{avg_spec_ophA}, but for L1688-SR2.}
     \label{avg_spec_top} 
\end{figure*}

\begin{figure*}[!ht]  
\centering
\includegraphics[width=0.98\textwidth]{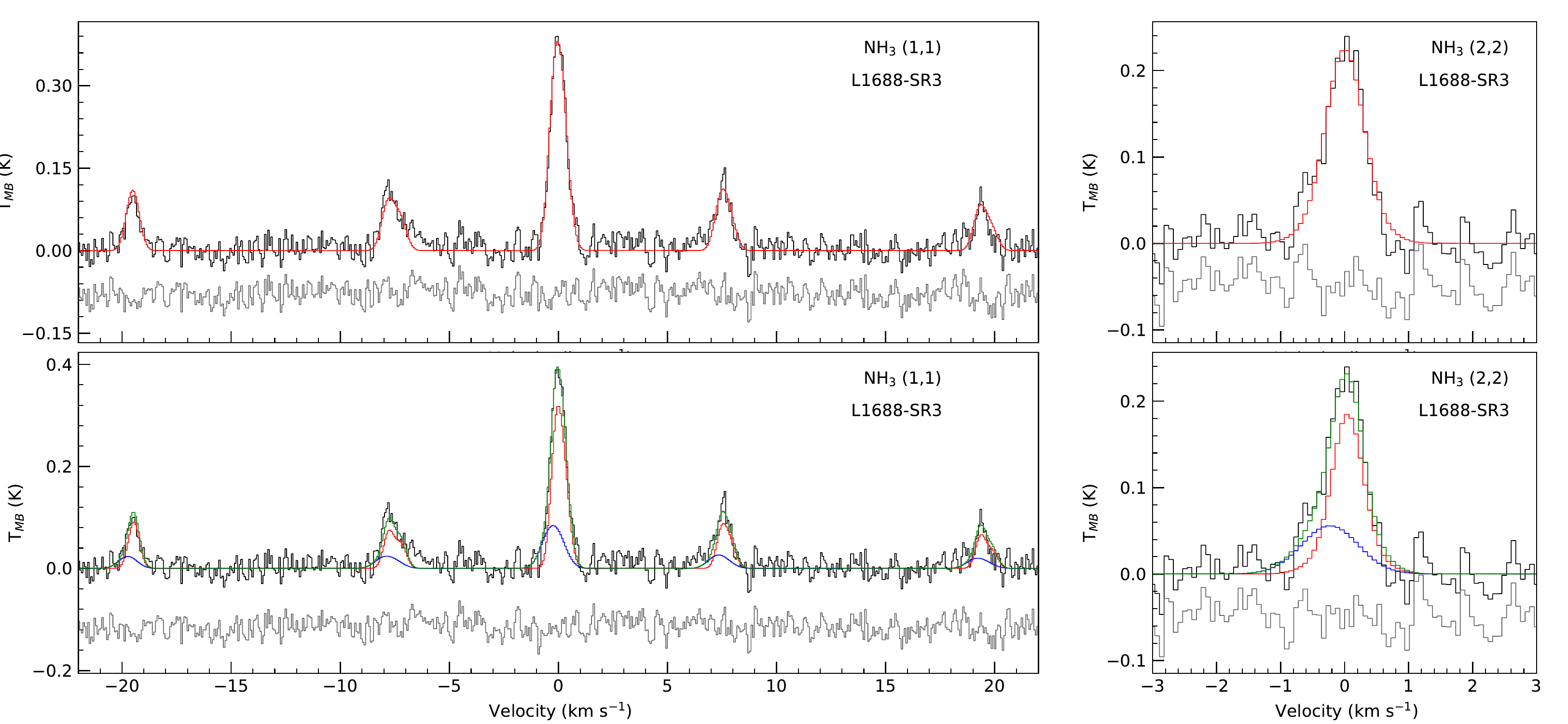}
  \caption{Same as Figure \ref{avg_spec_ophA}, but for L1688-SR3.}
     \label{avg_spec_west} 
\end{figure*}

Figures \ref{avg_spec_ophA} to \ref{avg_spec_west} show the average spectra in all of the other coherent cores studied here, with single- and two-component fits, similar to Oph-F in Figure \ref{avg_spec_ophF}. The kinetic temperature, velocity dispersion, centroid velocity and para-ammonia column densities determined from these single- and two-component fits are presented in Table \ref{tab_fit_para}. The rms noise in the spectra of each core is also shown in the same table.

\begin{table*}[ht]
\centering
\caption{Best fit parameters for single- and two-component fits}
\label{tab_fit_para}
\begin{tabular}{ccccccc|cc|cc}
\hline\hline            
Core & Component \tablefootmark{a} & \tk & \sig \tablefootmark{b} & $\rm v_{rel}$\tablefootmark{c} & $\log$(N(p-\amm)) & $\mathcal{M}_s$  & noise\tablefootmark{d} & $\Delta_{\rm AIC}$\tablefootmark{e} & $\rm T_{K, GAS}$ & $\rm \sigma_{v, GAS} $
 \\
 & & (K) & (\kms) & (\kms) &  (cm$^{-2}$) &  & (mK) & & (K) & (\kms)\\
 \hline

 & Single & 19.1(2) & 0.253(3) & 0.0\tablefootmark{f} & 13.86(2) & 0.9 &  &  & 19(1) & 0.23(1) \\ 
Oph-A & Narrow & 15.8(4) & 0.156(4) & -0.001(2) & 13.74(4) & 0.6 &  17 & 761 & & \\ 
 & Broad & 24.9(8) & 0.44(1) & -0.014(6) & -- \tablefootmark{g} & 1.5 &   & & & \\ 
\hline
 & Single & 11.72(9) & 0.177(1) & 0.0\tablefootmark{f} & 14.168(5) & 0.8 &  &  & 12.1(5) & 0.174(8)\\ 
Oph-C & Narrow & 10.13(9) & 0.1376(8) & -0.0015(4) & 14.088(6) & 0.6 &  9 & 6724 & & \\ 
 & Broad & 17.7(3) & 0.477(7) & -0.033(5) &14.13(3)  & 1.9&   & & & \\ 
\hline
 & Single & 9.9(2) & 0.104(1) & 0.0005(5) & 13.94(1) & 0.4&  &   & 10.2(4) & 0.116(7)\\ 
Oph-D & Narrow & 9.7(3) & 0.088(1) & -0.006(1) & 13.89(2)  & 0.3& 22, 21  & 383 & & \\ 
 & Broad & 12.0(1) & 0.39(2) & 0.14(2) &13.8(1) &  1.8 &  & & & \\ 
\hline
 & Single & 12.5(4) & 0.168(4) & 0.000(2) & 13.77(4) & 0.7 &  &  & 12.3(5) & 0.171(8)\\ 
Oph-E & Narrow & 11.7(3) & 0.126(2) & 0.023(2) & 13.73(3) & 0.5 & 24  & 3698 & & \\ 
 & Broad & 17.9(6) & 0.76(2) & -0.92(3) & -- \tablefootmark{g} & 3.0 &   & & & \\ 
\hline
 & Single & 13.9(1) & 0.174(2) & 0.0003(5) & 13.91(1)  & 0.7&   &  & 14.9(8) & 0.19(1)\\ 
Oph-F & Narrow & 12.6(1) & 0.128(1) & 0.006(1) & 13.81(1) & 0.5 & 9, 11  & 5244 & & \\ 
 & Broad & 19.1(4) & 0.55(1) & -0.144(9) &13.87(7) & 2.1&   &  & & \\ 
\hline
 & Single & 11.6(1) & 0.15(1) & 0.0008(5) & 14.0(9)  & 0.6&  &  & 12.4(4) & 0.16(1)\\ 
Oph-H-MM1 & Narrow & 11.0(1) & 0.133(1) & 0.012(1) & 13.971(9) & 0.6 & 16  & 1489 & & \\ 
 & Broad & 16.9(7) & 0.47(2) & -0.4(3) &13.99(8) & 1.9&   &  & & \\ 
\hline
 & Single & 12.1(2) & 0.137(2) & 0.005(1) & 13.79(2)  & 0.6&  &  & 12.2(2) & 0.160(8)\\ 
L1688-d10 & Narrow & 11.2(2) & 0.111(2) & 0.015(2) & 13.76(2) & 0.4 & 14  & 826 & & \\ 
 & Broad & 17.1(8) & 0.42(2) & -0.2(2) &13.6(2) & 1.7 &   & & & \\ 
\hline
 & Single & 10.6(2) & 0.135(1) & -0.0004(4) & 13.92(1)& 0.6  &  &  & 10.4(1) & 0.129(6)\\ 
L1688-d12 & Narrow & 9.4(6) & 0.123(4) & -0.0009(7) & 13.91(3) &0.6 &  16 & 30 &  & \\ 
 & Broad & 18.0(3) & 0.26(5) & 0.0(1) &13.6(4) & 1.0 &   & & & \\ 
\hline
 & Single & 12.4(3) & 0.149(3) & -0.002(2) & 13.81(3)  & 0.6 &  & & 12.5(5) & 0.14(1) \\ 
Oph-B3 & Narrow & 11.4(3) & 0.117(2) & -0.028(2) & 13.67(3) & 0.5 & 26, 28  & 1083 & & \\ 
 & Broad & 19.0(1) & 0.58(3) & 0.56(4) &13.8(2)  & 2.2 &   && & \\ 
\hline
 & Single & 13.5(4) & 0.209(6) & 0.000(1) & 13.75(5)  & 0.9&  &  & 14.2(5) & 0.18(2) \\ 
L1688-SR1 & Narrow & 10(1) & 0.102(5) & -0.033(4) & 13.3(1) &0.4 & 35  & 308 &  & \\ 
 & Broad & 18.0(1) & 0.43(2) & 0.11(2) &13.7(2) & 1.7 &   & & & \\ 
\hline
 & Single & 17.2(3) & 0.239(4) & 0.000(2) & 13.35(8) & 0.9 &  &  & 16.70(6) & 0.222(3)\\ 
L1688-SR2 & Narrow & 15.4(5) & 0.178(6) & -0.009(4) & 13.2(1) &0.7 & 11  & 164 &  & \\ 
 & Broad & 24.0(2) & 0.6(5) & 0.08(3) &13.0(4) & 2.0 &   & & & \\ 
\hline
 & Single & 24.6(6) & 0.295(8) & 0.0\tablefootmark{f} & -- \tablefootmark{g} & 0.9  &  & & -- \tablefootmark{h} & --  \tablefootmark{h}\\ 
L1688-SR3 & Narrow & 24(1) & 0.24(2) & 0.05(2) & -- \tablefootmark{g} & 0.8 & 18  & 12 & & \\ 
 & Broad & 27.0(5) & 0.41(7) & -0.2(2) & -- \tablefootmark{g} & 1.3 &   & & & \\ 
 \hline
 \hline
 &  &  &  &  &  &  &  &  \\ 
 \hline
 \hline
 & Single & 16.9(2) & 0.451(4) & 0.0002(0) & 13.79(2) & 1.8 &  &  &--  \tablefootmark{h} & --  \tablefootmark{h}\\ 
Ambient cloud \ \tablefootmark{i}& Broad-1 & 16.9(2) & 0.35(2) & -0.16(4) & 13.74(3) & 1.4 & 9  & 94 & & \\ 
 & Broad-2 & 17.0(5) & 0.35(3) & 0.45(7) & 13.2(3)  & 1.4 &   && & \\ 
\hline
\hline

\end{tabular}
\tablefoot{
Kinetic temperatures, velocity dispersions and p-\amm column densities, derived from single-component fits, and two-component fits, in the coherent cores and the ambient cloud \tablefootmark{i}. The values in parentheses represent the fit-determined error in the final decimal place of the corresponding parameter. These uncertainties do not include the calibration uncertainty, which is $\sim$10 \%. 
The noise level achieved in the average spectra are also shown.
The final two columns show the average kinetic temperature and velocity dispersions inside the cores, from GAS DR1 \citep{GASDR1} results (single-component fit).\\
\tablefoottext{a}{single-component fit, or the individual components of the} two-component fit. \\
\tablefoottext{b}{The channel response, $\sigma_{chan}$ (see Section \ref{sec_id_coh}), is not removed from the \sig value reported here (The contribution from $\sigma_{chan}$ is very small, changes only in the third decimal place in \sig).} \\
\tablefoottext{c}{Velocity from the fit. Since we align the spectrum in the core by the velocity at each pixel (determined from single-component fit), the velocities reported in this Table are relative to the mean velocity in the corresponding core.} \\
\tablefoottext{d}{Noise level estimated for both \amm (1,1) and (2,2). In cases where two values are reported, the noise in \amm (1,1) and (2,2) are not the same, and the values correspond to the noise in (1,1) and (2,2), respectively.} \\
\tablefoottext{e}{
Change in AIC, from 1-component fit to 2-component fit ($\rm \Delta_{ AIC} = AIC_{1-comp.} - AIC_{2-comp.}$).} \\
\tablefoottext{f}{Value and error smaller than 10$^{-4}$ \kms.} \\
\tablefoottext{g}{Excitation temperature could not be well-constrained from the fit (fit determined error > 30\%), and therefore, the column density estimate is not very reliable.} \\
\tablefoottext{h}{Velocity dispersion and temperature maps from GAS DR1 did not include any pixel inside SR3, or ambient cloud box.} \\
\tablefoottext{i}{Represents the average spectra in the ``ambient cloud box'' shown in Figure \ref{coh_cores}.} \\
}
\end{table*}

\end{document}